\documentclass[prl,twocolumn,preprintnumbers,amsmath,amssymb,superscriptaddress]{revtex4}
\usepackage{graphicx}
\usepackage{amsmath}
\usepackage{graphics}
\usepackage{amssymb}
\usepackage{layout}
\usepackage{verbatim}
\usepackage{amsfonts,epsfig}

\newcommand{\beq}{\begin{equation}}
\newcommand{\eeq}{\end{equation}}
\newcommand{\bea}{\begin{eqnarray}}
\newcommand{\eea}{\end{eqnarray}}

\begin{document}

\title{Absence of nonlocal resistance in microstructures of PbTe quantum wells }
\author{K. A. Kolwas}
\affiliation{Institute of Physics, Polish Academy of Sciences, al.~Lotnik{\'o}w 32/46, PL 02-668 Warszawa, Poland}
\author{G. Grabecki}
\affiliation{Institute of Physics, Polish Academy of Sciences, al.~Lotnik{\'o}w 32/46, PL 02-668 Warszawa, Poland}
\affiliation{Department of Mathematics and Natural Sciences, College of Sciences, UKSW, ul. Wóycickiego 1/3, PL-01-938 Warszawa, Poland}
\author{S. Trushkin}
\affiliation{Institute of Physics, Polish Academy of Sciences, al.~Lotnik{\'o}w 32/46, PL 02-668 Warszawa, Poland}
\author{J. Wr{\'o}bel}
\affiliation{Institute of Physics, Polish Academy of Sciences, al.~Lotnik{\'o}w 32/46, PL 02-668 Warszawa, Poland}
\author{M. Aleszkiewicz}
\affiliation{Institute of Physics, Polish Academy of Sciences, al.~Lotnik{\'o}w 32/46, PL 02-668 Warszawa, Poland}
\author{{\L}. Cywi{\'n}ski}
\affiliation{Institute of Physics, Polish Academy of Sciences, al.~Lotnik{\'o}w 32/46, PL 02-668 Warszawa, Poland}
\author{T. Dietl}
\affiliation{Institute of Physics, Polish Academy of Sciences, al.~Lotnik{\'o}w 32/46, PL 02-668 Warszawa, Poland}
\affiliation{Institute of Theoretical Physics, Faculty of Physics, University of Warsaw, ul. Ho{\.z}a 69, PL-00-681 Warszawa, Poland}
\author{G. Springholz}
\affiliation{Institut für Halbleiter-und-Festk{\"o}rperphysik, Johannes Kepler University, Altenbergerstr. 69, A-4040 Linz, Austria}
\author{G. Bauer}
\affiliation{Institut für Halbleiter-und-Festk{\"o}rperphysik, Johannes Kepler University, Altenbergerstr. 69, A-4040 Linz, Austria}
\date{\today }
\begin{abstract}
We report on experiments allowing to set an upper limit on the magnitude of the spin Hall effect and the conductance by edge channels in quantum wells of PbTe embedded between PbEuTe barriers. We reexamine previous data obtained for epitaxial microstructures of n-type PbSe and PbTe, in which pronounced nonlocal effects and reproducible magnetoresistance oscillations were found. Here we show that these effects are brought about by a quasi-periodic network of threading dislocations adjacent to the BaF$_{2}$ substrate, which give rise to a p-type interfacial layer and an associated parasitic parallel conductance. We then present results of transport measurements on microstructures of modulation doped PbTe/(Pb,Eu)Te:Bi heterostructures for which the influence of parasitic parallel conductance is minimized, and for which   quantum Hall transport had been observed, on similar samples, previously. These structures are of H-shaped geometry and they are patterned of 12 nm thick strained PbTe quantum wells embedded between Pb$_{0.92}$Eu$_{0.08}$Te barriers. The structures have different lateral sizes corresponding to both diffusive and ballistic electron transport in non-equivalent L valleys. For these structures no nonlocal resistance is detected confirming that PbTe is a trivial insulator. The magnitude of spin Hall angle $\gamma$ is estimated to be smaller than 0.02 for PbTe/PbEuTe microstructures in the diffusive regime. 
\end{abstract}


\maketitle

\section{I. Introduction}
Nonlocal resistance $R_{\rm nl}$ in multiterminal structures is defined as the ratio of voltage occurring in regions far from the nominal current path to the magnitude of the current. For example, in the case of standard Hall bar geometry [see figure 1 (inset)], $R_{\rm nl}$ is a ratio of the voltage measured between one pair of the Hall probes (e.g., 3-5) to the electric current flowing between the second pair (4-6). When the Hall bar is patterned of a uniform conductor then according to the classical electrodynamics, $R_{\rm nl}$ vanishes exponentially with the distance L between the two Hall arms according to
\beq
R_{\rm nl} = R_{\rm sq} \exp \left(-\pi\frac{L }{W} \right ) \,\, , 
\eeq
where $R_{\rm sq}$ is the sheet resistance, and $W$ is the width of the Hall bar \cite{vanderPauw_PTR58}. However, under certain conditions, the above formula ceases to be valid. For example, in the ballistic regime ($L$, $W \! < \! l_{e}$ - electron mean free path) much larger values of $R_{\rm nl}$ are detected when a fraction of the electron beam injected from one current contact enters directly into one of the voltage contacts \cite{Shepard_PRL92,Mihajlovic_PRL09}. Similarly, in the diffusive regime, the nonlocal effects are observed at low temperatures when the electron coherence length $L_{\Phi}$  is longer than $L$ \cite{Haucke_PRB90}.

Recently, nonlocal effects have been widely employed in studies of spin-polarized transport in metals and semiconductors. The spin accumulation caused by current injection from a ferromagnet to a paramagnet is now routinely sensed in non-local spin valves \cite{Jedema_Nature02,Lou_PRL06,Li_NC11}, in which the nonlocal voltage appears because the spin density diffuses also into the regions of the sample where no charge current flows, and spin-selective floating contacts are sensitive to this spin density. Another source of non-local resistance, important for this paper, is the spin Hall effect (SHE) appearing in Hall bar structures \cite{Dyakonov_JETP71,Hirsch_PRL99}. More precisely, a non-zero value of $R_{\rm nl}$ is brought about by the SHE and the inverse SHE, which conspire to generate a nonlocal signal: in one Hall arm, the charge current drives (via a spin-orbit interaction) a transverse spin current, and by the inverse process this spin current generates a transverse charge current in the other arm \cite{Hankiewicz_PRB04,Abanin_PRB09}. This second order (in spin-orbit coupling) effect is expected to appear in both ballistic and diffusive regimes, provided that the distance $L$ between Hall arms is shorter than the relevant spin diffusion length $L_{s}$. Recently, it has been observed for ballistic holes in nanostructures of HgTe \cite{Brune_NP10}.

The appearance of nonlocal resistances in a \emph{macroscopic} Hall structure requires the formation of robust conductive edge channels that can guide a current between the distant probes. It is well known that such states appear when a high-mobility two-dimensional electron gas is subjected to high magnetic fields \cite{Prange_QHE}. In this quantum Hall effect (QHE) regime, the conduction proceeds exclusively through the edge states extending along the structure perimeter, provided that the electron states in the interior are localized \cite{Halperin_PRB82,MacDonald_PRB84}. Electron motion along the edges occurs only in one direction, as the magnetic field breaks the time reversal symmetry, and the one-dimensional edge channels are chiral, i.e, only one direction of propagation is allowed at each edge. The edge channels may extend over macroscopic distances ($L\! >\! 1$ mm) because they are robust against backscattering. Indeed, they give rise to sizable nonlocal resistances observed in multiterminal macroscopic Hall structures \cite{McEuen_PRL90,Takaoka_SSC91,Dolgopolov_JETP91}. 

Recently, it has been revealed that a similar situation may occur even in the absence of a magnetic field ($B\! =\! 0$). This is possible in a class of narrow-gap semiconductors, the so-called topological insulators (TI), characterized by the presence of extended surface or edge states which lead to non-zero values of conductance even for the Fermi energy in the region of the bulk bang gap. A necessary condition for the existence of these TI gapless states is an inversion of conduction and valence bands caused by the influence of relativistic effects on the band structure \cite{Hasan_RMP10,Kane_Z2_PRL05,Bernevig_Science06}. In an analogy to the QHE, the charge transport through the TI edge states in two dimensional structures is called the quantum spin Hall effect (QSHE). In the QSHE, two counter-propagating helical edge states (i.e., with carriers of opposite spins propagating in opposite directions) are formed, going around the whole structure. Because typically there is no electron localization in the bulk at $B\! =\! 0$ in narrow gap semiconductors, and additionally, the interfacial space charge layers are often present, in order to observe a nonlocal resistance due to these states one has to deplete the sample by a gate. Bernevig et al.~\cite{Bernevig_Science06} theoretically predicted the existence of the QSHE in HgTe/Hg$_{1-x}$Cd$_{x}$Te quantum wells, and this prediction was experimentally confirmed by Konig et al.~\cite{Konig_Science07}. Subsequently, Roth et al.~\cite{Roth_Science09} studied mesoscopic Hall bars of the same wells and observed nonlocal resistances of magnitudes close to theoretically predicted quantized values. These results also suggested that in larger structures, the helical edge channels may be suppressed by potential fluctuations and/or inelastic scattering. On the other hand, Gusev et al.~\cite{Gusev_PRB11} studied macroscopic ($\sim \! 1$ mm) Hall bars of HgTe/HgCdTe QWs, and reported nonlocal resistances as high as 1 M$\Omega$, which were virtually temperature independent below 1 K. Clear evidences for helical edge modes were also found in InAs/GaSb quantum wells but nonlocal resistance measurements were not performed, as the bulk could not be depleted in this system \cite{Knez_PRL11}.

Recently, the existence of topologically protected gapless surface states has been predicted theoretically for Pb$_{1-x}$Sn$_{x}$Te with $x \! > \! 0.36$, in which the conduction and valence bands are inverted with respect to PbTe \cite{Hsieh_arXiv12}. These predictions were confirmed by ARPES measurements on band-inverted Pb$_{0.77}$Sn$_{0.23}$Se \cite{Dziawa_arXiv12} and Pb$_{0.65}$Sn$_{0.35}$Te \cite{Xu_arXiv12}.
Since these TI states exist only on certain crystal surfaces, this class of materials is referred to as topological crystalline insulators \cite{Fu_PRL11}. Furthermore, it has been theoretically demonstrated that PbTe/Pb$_{1-x}$Sn$_{x}$Te heterostructures, with a thin layer of one material embedded within the other, should have electronic properties analogous to thin layers of TIs for a certain range of Sn concentrations \cite{Buczko_PRB12}.

Surprisingly, however, there have been experimental data obtained by our group some time ago, showing pronounced nonlocal resistances and reproducible magnetoresistance oscillations in microstructures of PbTe and PbSe epilayers deposited on BaF$_{2}$ substrates (see figure 1)  \cite{Grabecki_SM97,Wrobel_APPA96}. Similar effects were also reported by others for wide quantum wells of PbTe \cite{Oswald_SSC97}. These findings were, of course, perplexing that time, and remain even more puzzling now, because PbTe and PbSe are expected to be trivial and not topological insulators \cite{Hsieh_arXiv12,Fu_PRB07}. 

In the first part of this work we present results of new careful transport measurements carried out on the previously \cite{Grabecki_SM97,Wrobel_APPA96} studied PbTe epilayers. Our findings demonstrate that the nonlocal resistance does not result from the edge channel transport, but originates from the presence of an additional conducting layer brought about by threading dislocations and/or the doping by the substrate material, in the interfacial region of the epilayers adjacent to the lattice mismatched BaF$_{2}$ substrate. The presence of these dislocations explains also the observation of anomalous reproducible magnetoresistance oscillations. In the second part, we present results obtained for PbTe modulation doped quantum wells in which the effect of leakage through the interfacial layer is minimized by a sufficiently thick Pb$_{1-x}$Eu$_{x}$Te buffer layer. Previously it had been shown \cite{Springholz_APL93,Olver_SSC94,Chitta_PRB05} that quantum Hall effect can be observed in such molecular beam epitaxy grown PbEuTe/PbTe/PbEuTe structures, i.e.~edge channel transport was demonstrated for applied magnetic fields. Furthermore, in narrow constrictions fabricated from such /PbTequantum wells embedded in PbEuTe barriers our group has reported precise zero-field conductance quantization in units of  of $2e^{2}/h$ \cite{Grabecki_PRB05,GrabeckI_JAP07}. We demonstrate that electron transport measurements on H-shaped microstructures do not show any nonlocal resistance in this case. Our data confirm, therefore, the trivial insulator character of PbTe. At the same time, they allow us to set an upper limit for the magnitude of the electrically detected spin Hall effect. We discuss these findings in the light of theoretical studies on the spin Hall effect in IV-VI semiconductors \cite{Murakami_PRL04,Dyrdal_EPL09}.

\begin{figure}[t]
\begin{center}
\includegraphics[width = 0.9\linewidth]{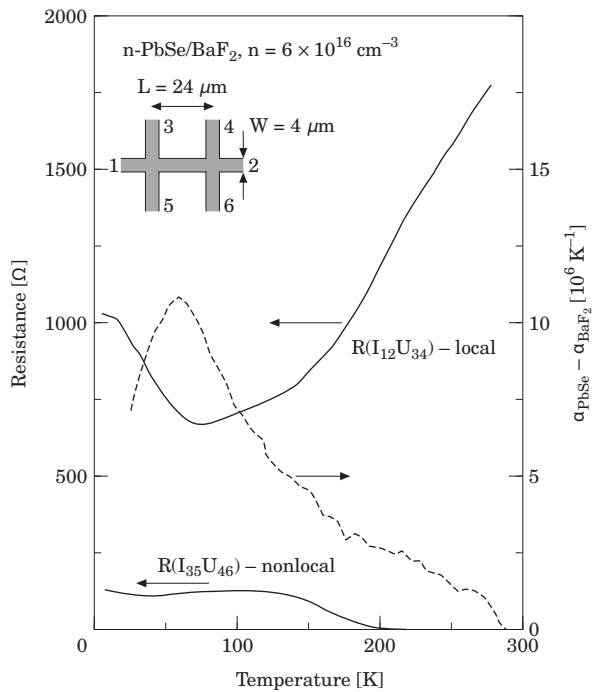}
\caption{(Color online) Local and nonlocal resistances measured as a function of temperature, at $B\! =\! 0$, in one PbSe microstructure deposited on BaF$_{2}$ substrate (adopted from Ref.~\cite{Grabecki_SM97}). }
\label{fig:1}
\end{center}
\end{figure}

Our paper is organized as follows: In section II we briefly review basic properties of PbTe. In section III we present results of the test measurements performed on one of the previous PbTe microstructures. Fabrication procedures and basic properties of the new microstructures, patterned of PbTe/Pb$_{1-x}$Eu$_{x}$Te quantum wells, are described in section IV. In Section V we present and discuss the results of transport measurements performed on the new structures. Section VI summarizes the present work.

\section{II. Basic properties of P\lowercase{b}T\lowercase{e}}  \label{sec:basic}
Similarly to HgTe, PbTe is a narrow gap semiconductor consisting of heavy elements, and its band structure is strongly modified by relativistic effects \cite{PbTe_review}. In particular, they account for an anomalous band gap increase with temperature and its decrease under pressure \cite{Tung_PR69,Schluter_PRB75,Gao_PRB08,Lusakowski_PRB11}. Although such a behaviour is similar to that of HgTe, current theory predicts that PbTe is rather a trivial insulator \cite{Hsieh_arXiv12,Fu_PRB07}, while strained HgTe (in which the energy gap is opened) is in fact a three-dimensional TI \cite{Brune_PRL11}.

There are also many other interesting differences between these two materials, especially regarding their spin properties. In particular, PbTe crystallizes in the rock-salt structure and has fourfold degenerate conduction and valence band extrema at the $L$ points of the Brillouin zone. Because the lattice possesses inversion symmetry, there is \emph{no} Dresselhaus spin-orbit term. Remarkably, PbTe is paraelectric and its static dielectric constant becomes huge at low temperatures, $\epsilon_{0} \! >\! 1000$ \cite{Bauer_ASSP83}. This implies strong dielectric screening, and a suppression of long-range parts of the Coulomb potentials of charged impurities. This makes PbTe a unique medium where the quantum lifetime becomes equal to the transport lifetime \cite{Dietl_JP78}. In such a case, the quantum ballistic regime is easily achievable in small structures \cite{Grabecki_PRB05}. For the same reason, the confinement potential in PbTe quantum wells preserves a rectangular shape, which strongly reduces the Rashba spin-orbit interaction. In view of these properties, even despite of the atomic spin-orbit interaction and considerable mixing of the spin "up" and "down" bands there are few mechanisms causing spin rotation in PbTe \cite{Jedrzejczak_PRB78}. This explains a long spin-orbit relaxation time in this material determined by weak localization \cite{Prinz_PRB99} and optical pumping experiments, $\tau_{so} \! \approx \! 2$ ns \cite{Kaufmann_PRB84}. Interestingly, PbTe was proposed as a suitable candidate for the observation of the SHE already some time ago \cite{Murakami_PRL04}. On the other hand, it was recently theoretically predicted \cite{Dyrdal_EPL09} that the magnitude of the intrinsic SHE in IV-VI compounds is proportional to the difference between the effective masses of electrons in the conduction band and holes in the valence bands, which is rather small, as these bands show nearly perfect particle-hole symmetry in PbTe. 

In order to fabricate the multiprobe PbTe structures suitable to study of the nonlocal effects, initial material in a form of thin epilayers or quantum wells should be employed. Because of the lack of a perfect lattice-matched substrate and resulting misfit dislocations, the electron mobility in PbTe epilayers is typically lower than in the corresponding bulk material. The most suitable substrate for PbTe growth by molecular beam epitaxy (MBE) is BaF$_{2}$ \cite{Springholz_03}. However, even in this case the layers contain a significant number of threading dislocations due to the 4.2\% lattice mismatch. Fortunately, the dislocation density decreases significantly with the layer thickness \cite{Springholz_APL96}. This is the origin of the observed strong dependence of the electron mobility of PbTe layers on their thicknesses \cite{Ueta_JCG97}. Importantly, the dislocation-induced defects act as p-type dopants. The density of dislocations in the vicinity of the substrate is so high that they lead to the formation of a $p$-type interfacial layer of a thickness $d \! \leq \! 0.5$  $\mu$m and high hole concentration, $p \! \sim \! 10^{19}$ cm$^{-3}$. Such a layer was studied in transport measurements on PbTe epilayers of different thickness \cite{Tranta88}. This $p$-type layer is also present in the case of Pb$_{1-x}$Eu$_{x}$Te grown on BaF$_{2}$ \cite{Grabecki_PE04}. Therefore, the PbTe epilayers which are nominally $n$-type in fact contain two conducting channels: the upper one is $n$-type and the lower one $p$-type. They are separated by a $p$-$n$ junction, as sketched in figure 2(a).

Additionally, the difference in thermal expansion coefficients between PbTe and BaF$_{2}$ generates thermal strains of the order of 0.16\% when the structure is cooled down to cryogenic temperatures \cite{Grabecki_PE04}. There are experimental evidence that this strain induces a movement of dislocations and produces additional defects in this rather elastically soft compound \cite{Zogg_PRB94}.

\begin{figure}[t]
\begin{center}
\includegraphics[width = 0.9\linewidth]{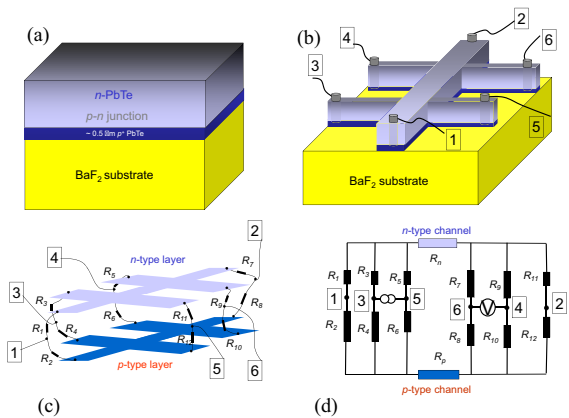}
\caption{(Color online) (a) Schematic cross-section of the $n$-PbTe epilayer showing the $p$-type interfacial layer close to the substrate. (b) Schematic view of a Hall bridge with indium diffusion at the contacts indicated. (c) Schematic picture of two conducting layers with resistances $R_{1}$...$R_{12}$ connecting six leads with n-type and p-type conducting layers, respectively. (d) The corresponding electric circuit serving to explain the presence of nonlocal effects. }
\label{fig:2}
\end{center}
\end{figure}

\section{III. Reexamination of results for P\lowercase{b}S\lowercase{e} and P\lowercase{b}T\lowercase{e} epilayers}  \label{sec:reexamination}

Some time ago, we studied electron transport properties of PbSe and PbTe microstructures patterned of $n$-type epilayers deposited on BaF$_{2}$ substrates \cite{Grabecki_SM97,Wrobel_APPA96}. The initial layer thicknesses were of the order of 1 $\mu$m, electron concentrations were in the range from $6\cdot 10^16$ to $6\cdot 10^{17}$ cm$^{-3}$ and Hall mobilities were between $1.2 \cdot 10^3$ and $2\cdot 10^5$ cm$^2$/Vs. Electron transport measurements revealed an unexpected nonlocal resistance appearing at low temperatures. Results for a PbSe epilayer are presented as an example in figure 1. 

Importantly, there was also another anomaly found in PbTe and PbSe epilayers, namely the occurrence of reproducible magnetoresistance oscillations \cite{Grabecki_JJAP95,Grabecki_SM97,Wrobel_APPA96}. Their amplitude was anomalously high and could not be explained within the standard theory of universal conductance fluctuations \cite{Altshuler_JETP86,Lee_PRB87}.

In order to reveal the origin of these anomalies, we have performed additional transport studies using three different PbTe samples, grown and processed according to the procedure described elsewhere (Refs.~\cite{Ueta_JCG97} and \cite{Grabecki_SM97,Wrobel_APPA96}, respectively). The layers have been patterned into a Hall bridges with dimensions $L\! =\!  1000$ $\mu$m and $W\! =\!  100$  $\mu$m [see figure 2(b)]. The first sample (E1) is a 3 $\mu$m thick $n$-type PbTe epilayer with the electron concentration $n_H  \! = \! 5.5 \cdot 10^{16}$ cm$^{-3}$ and a mobility $\mu_H \! = \! 1\cdot 10^6$ cm$^{2}$/Vs, according to Hall measurements at 4.2 K. The second one (E2) is a 4 $\mu$m thick $p$-type epilayer with $p_H \! = \! 4.5 \cdot 10^{17}$ cm$^{-3}$ and $\mu_H \! = \! 8.3\cdot 10^4$ cm$^{2}$/Vs. Finally, the third sample (E3) is an undoped 0.1 $\mu$m thick epilayer, chosen to provide information on the near-substrate interfacial layer. In this case, transport measurements at 4.2 K show a rather high hole concentration $p_H \! = \! 4 \cdot 10^{18}$ cm$^{-3}$ and low mobility, $\mu_H \! =\! 60$ cm$^{2}$/Vs. It alse has to be noted that for the $n$-type sample E1 we have used pure indium contacts, while for the $p$-type E2 and E3 the contacts have been made using gold chloride brown liquid \cite{Grabecki_PE04}.

\begin{figure}[t]
\begin{center}
\includegraphics[width = 0.9\linewidth]{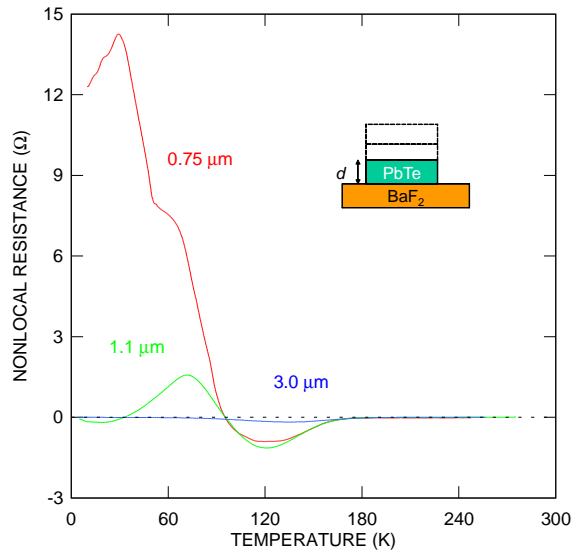}
\caption{(Color online) Nonlocal resistance as a function of temperature for sample E1 which has been consecutively thinned from 3 to 1.1 and further to 0.75 $\mu$m by chemical etching (as shown schematically in the inset).  }
\label{fig:3}
\end{center}
\end{figure}

\begin{figure}[t]
\begin{center}
\includegraphics[width = 0.9\linewidth]{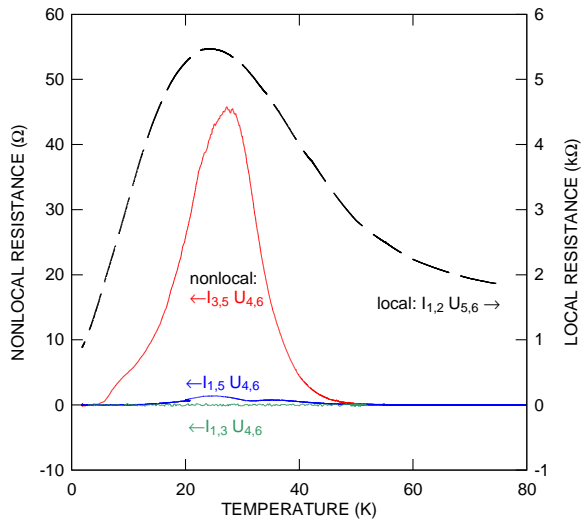}
\caption{(Color online) Local (right scale, dotted line) and nonlocal resistances (left scale, solid lines) as a function of temperature for the sample E1 thinned down to 0.75 $\mu$m (see figure 3). Nonlocal resistances are shown for three different contact probe configurations [see figure 2(b)].
}
\label{fig:4}
\end{center}
\end{figure}

We have performed a series of transport measurements for the consecutively thinned n-type structure E1 [figure 3 (inset)]. The thinning has been performed by wet chemical etching. We observe a strong reduction of the mobility values with decreasing layer thickness $d$, in agreement with results of the previous investigations \cite{Ueta_JCG97}. Simultaneously, we have found an enormous increase of the nonlocal resistance $R_{\rm nl}$, as shown in figure 3. For the initial layer, the magnitude of $R_{\rm nl}$ is rather small and detectable only between $80$ and $200$ K. For smaller $d$, its magnitude is higher by orders of magnitude. Interestingly, it also changes sign as a function of temperature but the behaviour of $R_{\rm nl}(T)$ is not reproducible after subsequent coolings from room temperature to $4.2$ K. Furthermore, the shape and magnitude of $R_{\rm nl}(T)$ vary if different probes are used for its determination, as depicted in figure 4 for $d\! =\! 0.75$ $\mu$m. Actually, we have found a correlation between the magnitude of $R_{\rm nl}(T)$ and the contact resistance. For the data presented in figure 4, the values of two probe resistances at $25$ K are $R_{13} \! =\! 1.8$ k$\Omega$, $R_{15} \! =\! 380$ k$\Omega$, $R_{14} \! = \! 15$ k$\Omega$ and $R_{15} \! = \!18$ k$\Omega$. As it is seen, $R_{\rm nl}(T)$ is nonzero when the current flows through the highly resistive contact no.~5, i.e., for the two measurement configurations $R_{35,46}$ and $R_{15,46}$, but vanishes for $R_{13,46}$. Additionally, these results directly demonstrate that the nonlocal effect observed cannot be the result of an edge channel conduction, because in that case one expects: $R_{13,46} + R_{15,46} \! = \! R_{35,46}$. 

We have also checked that the nonlocal effects are absent in the two $p$-type PbTe epilayers, E2 and E3. For the latter, which consists only of the interfacial defect layer, its resistance measured as a function of the magnetic field shows aperiodic fluctuations that are superimposed onto a monotonous background and are reproducible when the field is swept up and down (see figure 5). 

The above findings clearly suggest that the nonlocal effects in $n$-PbTe epilayers result from the presence of parallel conductance associated with the buried p-type conducting layer. The mechanism responsible for this effect is illustrated in  figure 2(b)-(d). It is well known that In forms a good contact to n-type PbTe \cite{Rogalski,Zimin_PE02,Chang_JPD80,Grabecki_JAP10}, but strongly diffuses into this material \cite{Chang_JPD80,Grabecki_JAP10}. Furthermore, under the metal spots, an increased number of dislocations is formed as a result of the associated thermal stress. From Ref.~\cite{Zimin_PE02} it is known that a single dislocation threading the $p$-$n$ junction in PbTe, reduces its resistance by about 1 G$\Omega$. Therefore, despite of the donor character of indium, it provides also a conducting path to the $p$-type layer. Since the $n$-type and $p$-type layers are partly shunted at the contacts, the Hall bridge structure can be represented by an equivalent resistor circuit depicted in Figs. 2(c,d). $R_1, R_2,\, ..., R_{12}$, represent resistances of the six In contacts to the $n$-type and $p$-type layers, respectively. The latter resistances are expected to be much higher than the former. For the circuit shown in figure 2(d), it is clear that $R_{\rm nl} \! = \! V/I$ is zero when the contact resistances are all equal or their values are proportional, for example $R_{3}/R_{5} \! =\! R_{4}/R_{6}$. However, in reality this cannot be expected because Indium diffusion is a random process. Therefore, in general, $R_{\rm nl} \! \neq \! 0$. It has to be noted that there is no substitute for In contacts which work well with $n$-type PbTe at low temperatures. For example, using pure lead gives highly resistive contacts \cite{Grabecki_JAP10}. 

The resistor-network model reproduces all features observed in our experiments. In particular, $R_{\rm nl}$ increases with the resistance of the $n$-type layer. Since it depends directly on the asymmetry of the contact resistances, it can show different signs. The observed poor reproducibility for subsequent cooling cycles can be explained by dislocation motion triggered by thermal stress. As expected, the non-local effect is absent in p-type samples E2 and E3, because in this case the interfacial layer is totally shunted to the rest of the epilayer. Turning to the reproducible magnetoresistance oscillations we note that considering the microscopic size of the sample E3, their magnitude is too large to be explained in terms of the standard theory of the universal conductance fluctuations \cite{Altshuler_JETP86,Lee_PRB87}. According to Ref.~\cite{Springholz_APL96}, the misfit dislocation density in PbTe epilayer of $d\! =\! 100$ nm is of the order of $10^9$ cm$^{-2}$. Possibly, some quasi-periodic arrangement of these dislocations would give rise to the Altshuler-Aronov-Spivak interferences \cite{Beenakker}, in a way similar to large arrays of mesoscopic rings \cite{Pannetier_PRL84}. In order to check this possibility, we have performed Fast Fourier transformation (FFT) of the fluctuation pattern. The result is shown in inset to figure 5. To estimate the dimensions of the interference orbits $A$, the fluctuation frequency scale has been multiplied by half of the magnetic flux quantum ($h/2e$). Surprisingly, $A \! \ll \! 10^{-9}$  cm$^{-2}$, which means that the orbit sizes are significantly smaller than the average distance between the dislocations. Therefore, it is impossible that the interferences are caused by their quasi-periodic arrangement. Moreover, the orbit diameters are smaller than 100 nm, which is almost comparable to the diameters of the dislocation cores imaged by the high-resolution electron microscopy \cite{Springholz_03}. For such small orbits, one interesting explanation may be found in Ref.~\cite{Figielski_pssb00}, where similar conductance fluctuations have been observed in lattice mismatched III-V heterostructures. They were explained in terms of the Altshuler-Aronov-Spivak effect occurring when the electron waves encircle the Read cylinders of the dislocations. 

To summarize this section: we have found a new mechanism leading to a spurious nonlocal resistance in semiconductor microstructures. It appears if there is a parallel conductance channel and values of contact resistances vary among the contacts. 

\begin{figure}[t]
\begin{center}
\includegraphics[width = 0.9\linewidth]{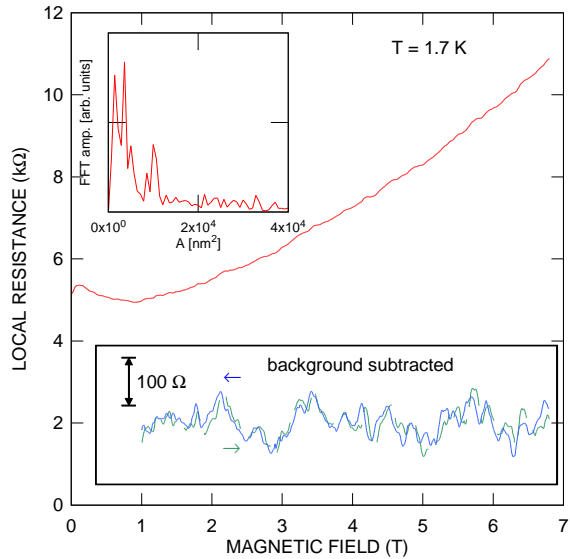}
\caption{(Color online) Local resistance as a function of magnetic field for the sample E3. Lower inset shows reproducible resistance fluctuations superimposed on the background. Upper inset shows FFT spectrum of the fluctuations measured in the magnetic field range from 0 to 3 T.
}
\label{fig:5}
\end{center}
\end{figure}

\section{IV. Fabrication of P\lowercase{b}T\lowercase{e}/P\lowercase{b}$_{1-x}$E\lowercase{u}$_{x}$T\lowercase{e} microsctructures} \label{sec:fabrication}

In order to search for edge channel transport and the spin Hall effect, we have fabricated a series of microstructures of PbTe quantum wells embedded in PbEuTe layers grown by MBE on BaF$_{2}$ substrates. They contain a 12 nm wide $n$-type PbTe QW embedded between Pb$_{0.92}$Eu$_{0.08}$Te barriers \cite{Springholz_03}. In order to introduce electrons into the quantum well, modulation doping by Bi donors ($N_D \! \approx \!  3\cdot 10^{18}$ cm$^{-3}$) is employed with an undoped 2 nm wide Pb$_{0.92}$Eu$_{0.08}$Te spacer layer separating the quantum well and the doping layer. A 2.8 $\mu$m thick undoped Pb$_{0.92}$Eu$_{0.08}$Te buffer layer serves to separate the structure from the substrate and, thus, from the $p$-type interfacial layer. The energy bandgap of the barrier and buffer is as large as 550 meV at helium temperatures \cite{Yuan_PRB93}. In this way, the barrier height in the $p-n$ junction is strongly increased.

Because of the [111] growth direction, the fourfold $L$-valley degeneracy of the conduction band in PbTe QW is lifted by the quantum confinement, so that the ground-state two-dimensional (2D) subband is formed of the single valley with its main axis parallel to the growth direction \cite{Grabecki_PRB05} The lowest states formed by the three remaining valleys obliquely oriented to the growth directions, have a higher energy of 35 meV above the ground state for this QW thickness.

\begin{figure}[t]
\begin{center}
\includegraphics[width = 0.9\linewidth]{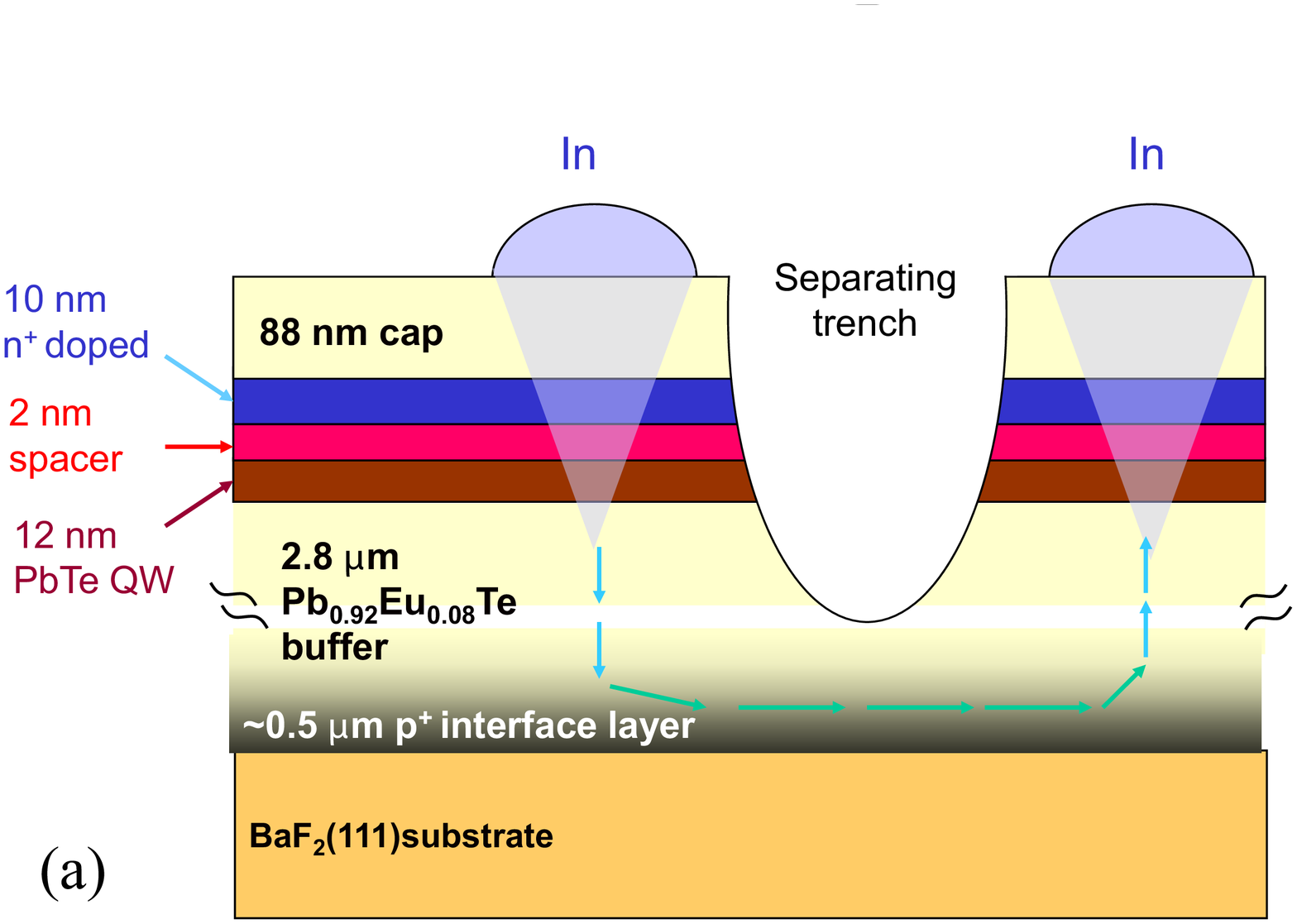}
\includegraphics[width = 0.9\linewidth]{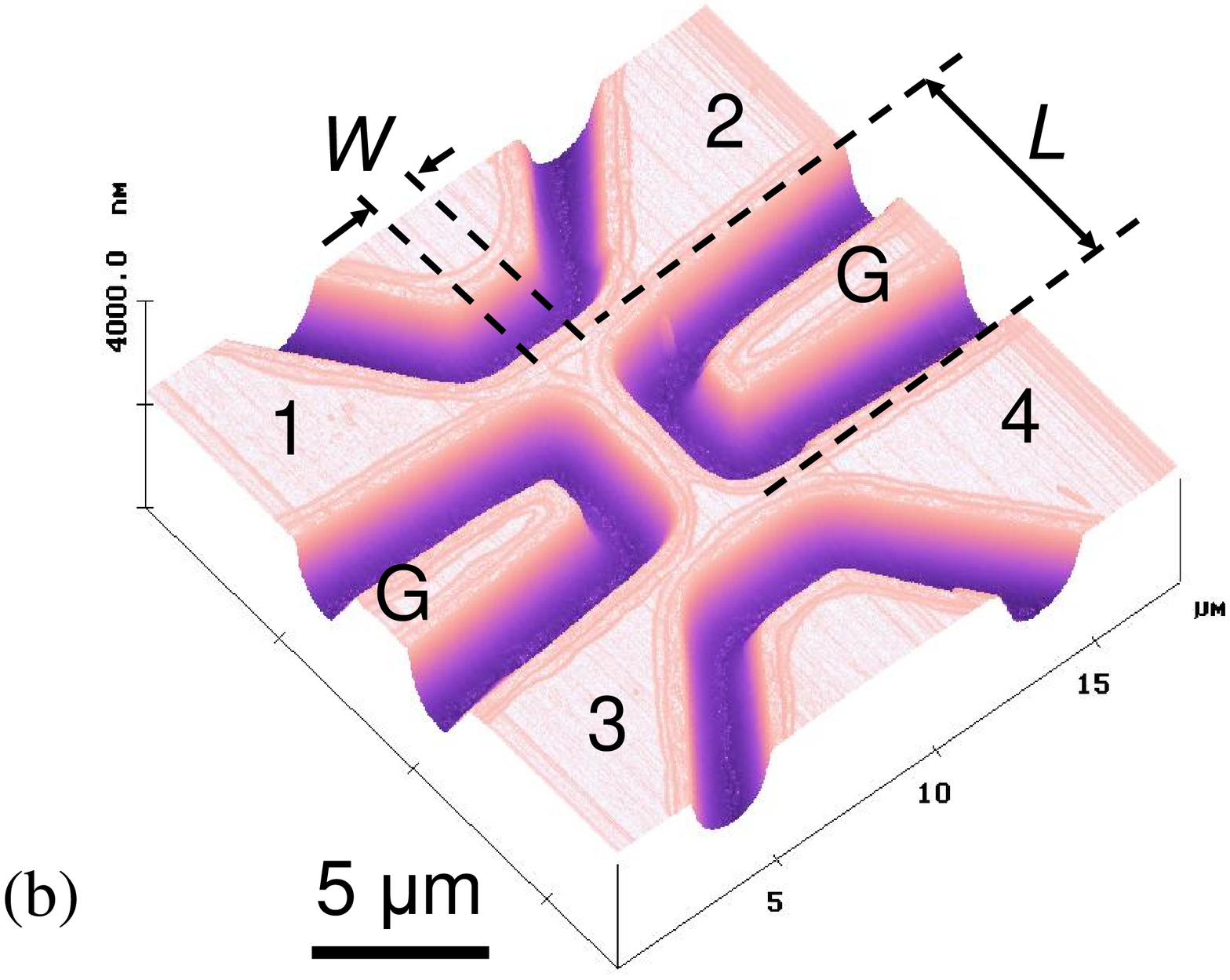}
\caption{(Color online) (a) Layout of PbTe/Pb$_{0.92}$Eu$_{0.08}$Te layer grown on BaF$_2$ substrate. Indium diffusion under the contact spot is schematically indicated. (b) Atomic force microscopy image of a H-shaped structure with side gates (G) patterned of PbTe quantum well. 
}
\label{fig:6}
\end{center}
\end{figure}

Two series of microstructures have been patterned from two wafers (A and B) with different sheet electron concentrations $n$ and electron mobilities $\mu$. For the wafer A: $n \! =\! 8\cdot 10^{12}$  cm$^{-2}$ and $\mu \! = \! 5\cdot 10^4$ cm$^2$/Vs, for the wafer B: $n \! =\! 1.5\cdot 10^{12}$ cm$^{-2}$ and $\mu  \! = \! 11\cdot 10^4$ cm$^2$/Vs at 4.2 K. These values allow us to estimate the electron diffusion constant $D$, electron mean free path $l_e$, and the spin diffusion length $L_s$. For the wafer A: $D \! \approx \! 0.8 \cdot 10^4$ cm$^2$/s, $l_e \! \approx \! 1.2$ $\mu$m, and  $L_{s} = \sqrt{D\tau_{s}} \approx 40$ $\mu$m. For layer B: $D \! \approx \! 0.33 \cdot 10^4$ cm$^2$/s, $l_e \! \approx \! 1.1$ $\mu$m, and  $L_{s} = \sqrt{D\tau_{s}} \approx 26$ $\mu$m.

The microstructures have been patterned in the form of H-shaped structures [see figure 6(b)], which are widely used in studies of the nonlocal resistance \cite{Hankiewicz_PRB04,Brune_NP10,Roth_Science09}. We have employed electron-beam lithography and wet chemical etching, in the same way as reported in our earlier work \cite{Grabecki_PRB05}. The etched trenches that define the H structures have a depth of about 0.9 $\mu$m, so they do not reach the substrate and get only to the buffer layer. Five microstructures of different lengths $L$ and widths $W$ have been used for the present measurements, as listed in Table~1.

\begin{table}
\begin{tabular}{ccccc}
Wafer 	& 	Structure No.	&	 $L\times W$ [$\mu$m$^2$]	&	$n_{\rm{2D}}$ [cm$^2$]	&	$l_{e}$ [$\mu$m] \\
\hline
A				&		A1						&		$10.5 \times 4$					&	$8\cdot 10^{12}$						&	$1.2$ \\	
A				&		A2						&		$7.5 \times 2$					&	$8\cdot 10^{12}$						&	$1.2$ \\	
A				&		A3						&		$6 \times 1$						&	$8\cdot 10^{12}$						&	$1.2$ \\	
B				&		B1						&		$45 \times 7$						&	$1.5\cdot 10^{12}$					&	$1.1$ \\	
B				&		B2						&		$50 \times 8$						&	$1.5\cdot 10^{12}$					&	$1.1$ \\
\end{tabular}
\end{table}

Each structure has been equipped with a pair of side gates for tuning the structure's conductance. The tuning is quite effective mainly because of the huge dielectric constant of PbTe which leads to rather high values of the gate capacitance. All leads, either to the structure or to the side electrodes, are connected to 300 $\mu$m wide contact pads. Gold wires are thermally soldered by indium to the upper surfaces of these pads. We notice here that any residual In diffusion across the buffer layer also causes a finite gate resistance [see scheme in figure 6(a)]. Therefore, in contrast to the old structures we have a possibility to directly estimate the parallel resistance by measuring the gate current in the different samples. All measurements are performed by both AC (at 7 Hz) and DC methods. In this way, we can control the magnitude of parasitic signals arising from capacitive coupling.

\section{V. Results and discussion}  \label{sec:results}
As a first step, we have checked whether the microstructurization process alters the initial properties of the PbTe QWs and, in particular, whether the mean free path remains sufficiently long to observe signatures of ballistic transport in the smallest structures (see table 1). The results displayed in figure 7 indeed demonstrate the appearance of a cross-over from the diffusive to ballistic regime with reduced sample dimensions. In particular, we observe a negative magnetoresistance, the magnitude of which grows when $W$ is reduced. If $W \! \leq \! l_e$, the resistance increases due to a contribution from electron backscattering at the entrance to the Hall arm of the H-structure. However, the backscattering is suppressed in a perpendicular magnetic field because electrons can enter the constriction owing to the formation of skipping orbits guided along the wire edges \cite{Beenakker}. The observation of the ballistic effect confirms that the electron mean free path $l_e$ has not been diminished by the microstructurization process. Since the temperature dependence of this magnetoresistance is rather very weak up to 30 K, it cannot be explained by weak-localization. The visible asymmetry of this magnetoresistance results from a contribution from the Hall effect. We have checked that the Onsager relations $R_{ij,kl}(B) = R_{kl,ij} (-B)$ are obeyed.

\begin{figure}[t]
\begin{center}
\includegraphics[width = 0.9\linewidth]{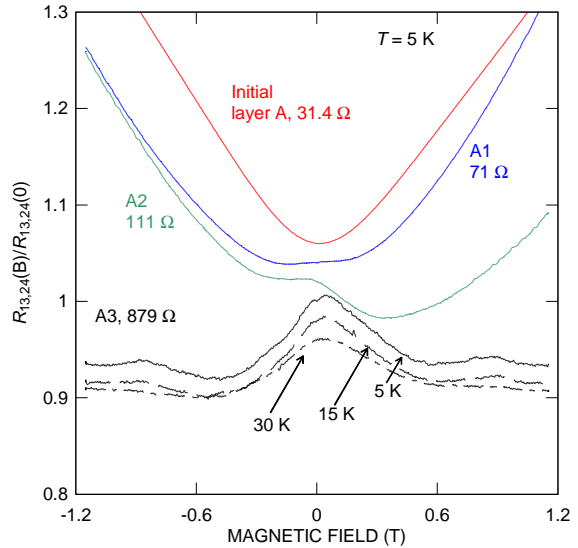}
\caption{(Color online) Relative magnetoresistance in the three H-structures, compared with the result for the macroscopic layer. The results have been obtained at $V_g  \! =\! 0$ and $T \! = \! 5$ K. Subsequent curves are labeled by the H-structure number and their resistances at $B\! =\! 0$. For the structure A3, the data for two additional temperatures are also shown. All curves are shifted vertically by $0.02$ to prevent overlap.
}
\label{fig:7}
\end{center}
\end{figure}

\begin{figure}[t]
\begin{center}
\includegraphics[width = 0.9\linewidth]{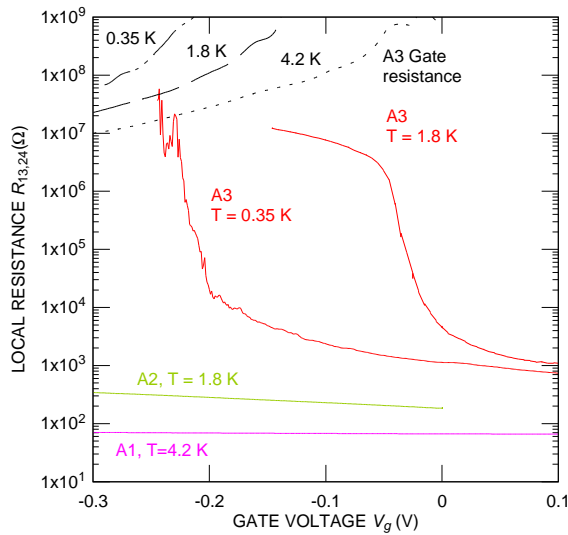}
\caption{(Color online) Local resistance as a function of the gate voltage for the three H-structures patterned from layer A. For structure A3, the data for two different temperatures are shown. All the measurements have been performed for the current 100 nA. Three upper curves represent the resistance measured between the side gate electrodes in the structure A3. 
}
\label{fig:8}
\end{center}
\end{figure}

\begin{figure}[t]
\begin{center}
\includegraphics[width = 0.9\linewidth]{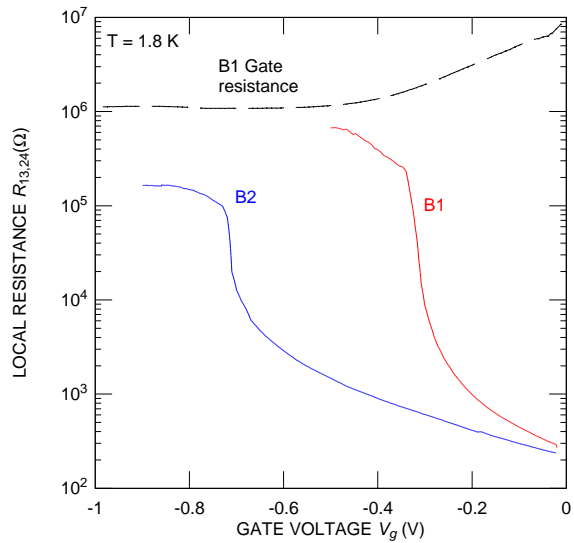}
\caption{(Color online) Local resistance as a function of the gate voltage for two H-structures patterned of layer B. The measurements have been performed for the current 100 nA. The upper curve represents the resistance measured between the side gate electrodes in the structure B1.
}
\label{fig:9}
\end{center}
\end{figure}

In order to search for edge channel transport one has to reach the depletion regime, in which the bulk is insulating. For structures patterned of the wafers A and B, the local resistances are presented in Figs.~8 and 9 as a function of the gate voltage $V_g$. A negative gate voltage increases the resistances, up to a certain level at which they saturate. Because of the high value of the initial electron density in structures patterned of wafer A, the depletion is achieved only for the smallest structure A3. For the structures A1 and A2 one observes only a minor increase of the resistance in the range up to $V_g \! = \!  1$ V. A difference in the threshold voltage $V_{\rm{th}}$ is found for sample A3 at 1.8 K and 0.35 K - which results from variations in $V_{\rm{th}}$ for subsequent cooling cycles. Importantly, the resistance of our smallest microstructure (A3) upon depletion shows saturation at a level as high as $10^7$ $\Omega$ at 0.35 K. This is the same order of magnitude as reported for the microstructure of a narrow HgTe QW ($d \! =\! 5.5$ nm), which is a trivial insulator, but significantly higher than the values $R \! <\! 10^5$ $\Omega$ found for microstructures of HgTe QWs in the TI regime ($d \!=\! 7.3$ nm) \cite{Konig_Science07}. Furthermore, according to magnetoresistance measurements presented above, electron transport in this microstructure exhibits ballistic features indicating that the channel width is shorter that the mean free path, also for inelastic collisions. We conclude that our data do not provide evidence for the presence of conducting edge states in microstructures of PbTe QWs. The three upper curves in figure 8 show the gate resistances for three different temperatures, plotted as a function of $V_g$. These are significantly higher than resistances of the structures measured in the whole range of $V_g$. This result clearly indicates that the contribution from the transport through the $p$-type interfacial layer is rather small. 

For the two relatively large microstructures patterned of wafer B, owing to their low electron density (figure 9), we observe the depletion in the studied range of $V_g$, as shown in figure 9. In this case, the resistance saturates at a relatively low value, not exceeding $10^6$ $\Omega$. Since, however, the measured gate resistance is of the same order, the resistance saturation is to be rather assigned to a leakage via the interfacial layer than to the presence of edge conducting channels.
 
\begin{figure}[t]
\begin{center}
\includegraphics[width = 0.9\linewidth]{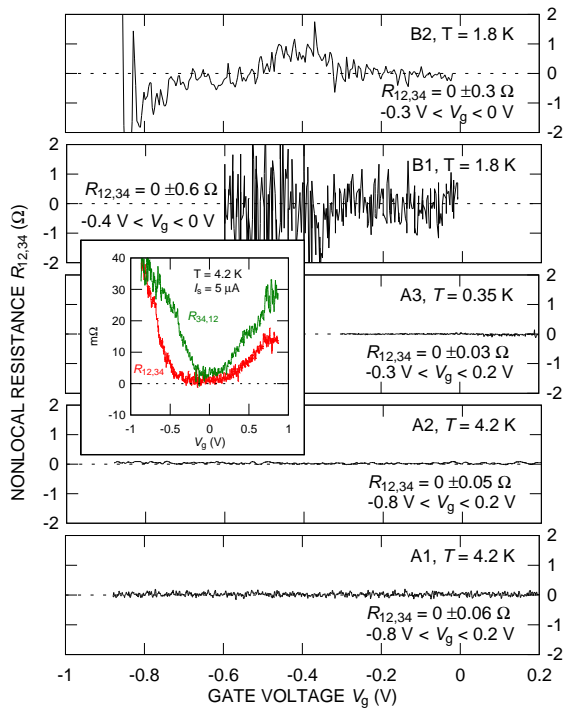}
\caption{(Color online) Nonlocal resistances at B=0 for the same H-shaped structures as in Figs.~8 and 9 for a current of 0.1 $\mu$A (the linear regime). Inset shows data for the A3 structure for a much higher sample current of 5 $\mu$A which is in the non-linear regime). The two solid lines correspond to the measurements with mutually exchanged voltage and current contacts. 
}
\label{fig:10}
\end{center}
\end{figure}

As emphasized in the Introduction, particularly direct information on the edge channel transport is provided by non-local resistance measurements. Furthermore, such studies allow to assess the magnitude of the spin Hall effect when the Fermi energy resides in the region of the bulk energy bands. We have found that for both groups of microstructures the nonlocal signals, $R_{\rm nl} \! = \!  R_{12,34}$, are below the noise level, as shown in figure 10. The corresponding upper limits determined for particular samples are: B2: $R_{12,34} \! = \! 0 \pm 0.3$ $\Omega$,  $-0.3 < V_g < 0$ V, B1: $R_{12,34} \! = \! 0 \pm 0.6$ $\Omega$,  $-0.4 < V_g < 0$ V, A3: $R_{12,34} \! =\! 0 \pm 0.03$ $\Omega$,  $-0.3 < V_g < 0.2$ V. Similarly, for the structures A1 and A2, which cannot be depleted, we have obtained $R_{12,34} \! =\! 0 \pm 0.06$ $\Omega$ and $0 \pm 0.06$ $\Omega$, respectively, within a range  $-0.8 < V_g < 0.2$ V. For these measurements we use small currents, $I_s \! \leq \! 200$ nA, i.e. within the linear regime. The range of applied $V_g$'s corresponds to values of local the resistance $R_{13,24}$ at least 100 times smaller than the gate resistance. 

It has to be noted, however, that in the geometry of the present system [figure 6(b)] the depletion is not uniform over the entire sample volume. In particular, we have checked by two-probe resistance measurements vs $V_g$ that the central wire is depleted prior to the two Hall arms. Therefore $R_{\rm nl}$ will be additionally reduced by the bulk resistance $R_{B}$. If the studied system was a topological insulator, $R_{\rm nl}$ could be estimated employing the Landauer-B{\"u}tiker formalism \cite{Roth_Science09}, taking into account the shunting resistance of the central wires, $R_{B}$:
\beq
R_{\rm{nl}} = \frac{R_{0}}{2} \frac{R^{2}_{B}}{(R_{0}+R_{B})(R_{0}+2R_{B})} \,\, ,
\eeq
where $R_0 \! =\! h/n_{v}e^{2}$ and $1 \leq n_v \leq 4$ is the number of valleys contributing to the edge channel transport. We have found that for all three structures exhibiting the depletion (see Figs.~8 and 9), $R_{B}$ at $V_g \! \approx \! V_{\rm{th}}$ is always larger than 1000 $\Omega$. Even in this limit, equation (2) would imply $R_{\rm nl}  \! \approx \! 31$ $\Omega$ for $n_v \! = \! 1$, which is two or three orders of magnitude greater than the upper limit of our experimental values. This evaluation points to the lack of conducting edge channels in our structures, which is consistent with the ordinary insulator behaviour of PbTe \cite{Hsieh_arXiv12,Fu_PRB07}.

The above results also suggest that the magnitude of the spin Hall effect is rather small in the PbTe QWs. Abanin et al \cite{Abanin_PRB09} derived an expression for the nonlocal resistance $R^{\rm{SH}}_{\rm{nl}}$  generated by extrinsic SHE in two-dimensional structures of a shape similar to ours,
\beq
R^{\rm{SH}}_{\rm{nl}} = 0.5 \gamma^{2}  R_{\rm{sq}}  \frac{W}{L_{s}}\exp \left(- \frac{L}{L_{s}} \right ) \,\, ,
\eeq 
where $\gamma$ denotes the spin Hall angle defined as the ratio of spin and charge conductivities. This equation is valid for the diffusive regime, i.e., $l_e \! \ll \! W$. For the two largest structures, B1 and B2, at $V_g$ close to $V_{\rm{th}}$ (see figure 9) we obtain an estimate for the spin-Hall angle $\gamma \! \leq \! 0.02$.

Unfortunately, the experimental resolution cannot be improved by increasing the magnitude of the current. An example is shown in the inset to figure 10. For $V_g \neq 0$, the value of $R_{\rm nl}$ determined at currents of the order of 5 $\mu$A, i.e. in the nonlinear regime, depends on the current magnitude and does not obey the Onsager relations, $R_{ij,kl} (B) \neq R_{kl,ij}(-B)$. This behaviour is presumably related to the appearance of a parallel conductance under the nonlinear conditions. Interestingly, an increase in the nonlocal signal is also detected for the H-structures in strong magnetic fields perpendicular to the structure plane. In figure 11, local and nonlocal resistances of structure A3 in the magnetic field B up to 8 T are presented. The local response shows a large positive magnetoresistance with superimposed Shubnikov-de Haas oscillations of electrons occupying several electric subbands in the PbTe QW. Three phenomena conspire to make the zero resistance states and, thus, nonlocal resistance specific to the quantum Hall effect, invisible in these structures: (i) presence of overlapping Landau levels due to the occupation of several electric subbands; (ii) reduced electron localization caused by the huge static dielectric constant; (iii)  presence of a residual parallel conductance, the relative importance of which increases with the magnetic field, since the contribution of high mobility QW electrons vanishes as $(\mu B)^{-2}$ at $\mu B\! >\! 1$. The parallel conductance by low mobility carriers accounts also for a large positive magnetoresistance. Instead of the nonlocal conductance associated with the quantum Hall effect, we observe a small nonlocal signal in the form of magnetoresistance fluctuations, fully reproducible when the magnetic field is swept up and down. The fluctuation amplitude increases with the magnetic field, but even in 8 T its magnitude is still in the $\sim \!0.1$ $\Omega$ range. In analogy to the results presented in Sec.~III, we assign this effect to the Altshuler-Aronov-Spivak interferences occurring in the misfit dislocation layer in the vicinity of BaF$_2$ substrate. FFT spectrum of the fluctuation pattern is shown in inset to figure 11. Note that the distribution of the interference orbit areas is considerably wider than in the case of the sample E3 [see figure 5(inset)]. Probably, the possible dislocation types and configurations and their densities are slightly different for the defect layer occurring in the elastically harder Pb$_{0.92}$Eu$_{0.08}$Te alloy, with respect to that in pure PbTe.

\begin{figure}[t]
\begin{center}
\includegraphics[width = 0.9\linewidth]{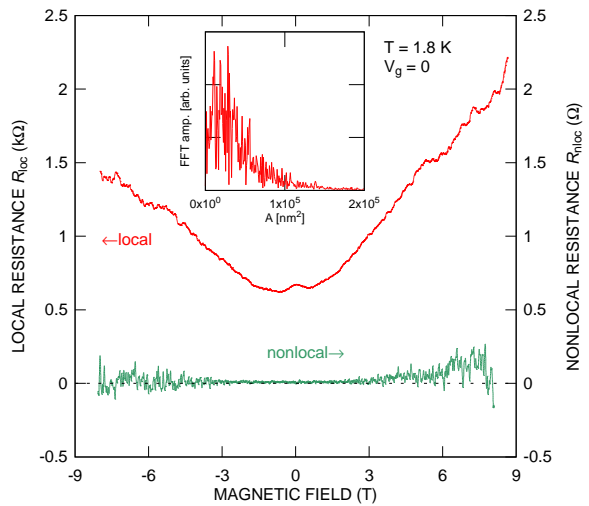}
\caption{(Color online) Local and nonlocal resistances as a function of magnetic field in microstructure A3. Note different vertical scales for the two curves. Inset shows FFT spectrum of the fluctuations measured in the magnetic field range from 0 to 3 T.
}
\label{fig:11}
\end{center}
\end{figure}

\section{VI. Summary} \label{sec:summary}
We have found that the previously observed nonlocal effects in epitaxial PbTe layers originate from parallel conductance by an interfacial layer close to the substrate, and asymmetries in the contact resistances. According to the results presented in this paper, the conducting holes are provided by misfit dislocations, whose peculiar electric properties explain also the presence of reproducible magnetoresistance oscillations. We have fabricated new PbTe/PbEuTe microstructures, in which contributions from these effects are minimized. These structures are suitable to search for signatures of the edge channel transport and/or of the spin Hall effect. No evidence for edge conductance has been found, which verifies experimentally the trivial insulator character of PbTe/Pb$_{0.92}$Eu$_{0.08}$Te quantum wells. We have also shown that despite strong relativistic effects brought about by the heaviness of the constituting elements, the magnitude of the spin Hall effect of electrons in PbTe remains below the present detection limit, presumably because the Dresselhaus and Rashba effects are absent in the studied PbTe quantum wells. The methodology we have developed makes it possible to test manifestations of the quantum and ordinary spin Hall effects in IV-VI compounds in both ballistic and diffusive regimes.
 
\section*{Acknowledgements} 
This work was supported by Ministry of Science (Poland) under Grants No.~1247/B/H03/2008/35 and No.~IP2010 006070 (Iuventus Plus).


\begin{thebibliography}{69}
\expandafter\ifx\csname natexlab\endcsname\relax\def\natexlab#1{#1}\fi
\expandafter\ifx\csname bibnamefont\endcsname\relax
  \def\bibnamefont#1{#1}\fi
\expandafter\ifx\csname bibfnamefont\endcsname\relax
  \def\bibfnamefont#1{#1}\fi
\expandafter\ifx\csname citenamefont\endcsname\relax
  \def\citenamefont#1{#1}\fi
\expandafter\ifx\csname url\endcsname\relax
  \def\url#1{\texttt{#1}}\fi
\expandafter\ifx\csname urlprefix\endcsname\relax\def\urlprefix{URL }\fi
\providecommand{\bibinfo}[2]{#2}
\providecommand{\eprint}[2][]{\url{#2}}

\bibitem[{\citenamefont{Van Der~Pauw}(1958)}]{vanderPauw_PTR58}
\bibinfo{author}{\bibfnamefont{L.~J.} \bibnamefont{Van Der~Pauw}},
  \bibinfo{journal}{Philips Tech. Rev.} \textbf{\bibinfo{volume}{20}},
  \bibinfo{pages}{220} (\bibinfo{year}{1958}).

\bibitem[{\citenamefont{Shepard et~al.}(1992)\citenamefont{Shepard, Roukes, and
  Van~der Gaag}}]{Shepard_PRL92}
\bibinfo{author}{\bibfnamefont{K.~L.} \bibnamefont{Shepard}},
  \bibinfo{author}{\bibfnamefont{M.~L.} \bibnamefont{Roukes}},
  \bibnamefont{and} \bibinfo{author}{\bibfnamefont{B.~P.} \bibnamefont{Van~der
  Gaag}}, \bibinfo{journal}{Phys.\ Rev.\ Lett.} \textbf{\bibinfo{volume}{68}},
  \bibinfo{pages}{2660} (\bibinfo{year}{1992}).

\bibitem[{\citenamefont{Mihajlovic et~al.}(2009)\citenamefont{Mihajlovic,
  Pearson, Garcia, Bader, and Hoffmann}}]{Mihajlovic_PRL09}
\bibinfo{author}{\bibfnamefont{G.}~\bibnamefont{Mihajlovic}},
  \bibinfo{author}{\bibfnamefont{J.~E.} \bibnamefont{Pearson}},
  \bibinfo{author}{\bibfnamefont{M.~A.} \bibnamefont{Garcia}},
  \bibinfo{author}{\bibfnamefont{S.~D.} \bibnamefont{Bader}}, \bibnamefont{and}
  \bibinfo{author}{\bibfnamefont{A.}~\bibnamefont{Hoffmann}},
  \bibinfo{journal}{Phys.\ Rev.\ Lett.} \textbf{\bibinfo{volume}{103}},
  \bibinfo{pages}{166601} (\bibinfo{year}{2009}).

\bibitem[{\citenamefont{Haucke et~al.}(1990)\citenamefont{Haucke, Washburn,
  Benoit, Umbach, and Webb}}]{Haucke_PRB90}
\bibinfo{author}{\bibfnamefont{H.}~\bibnamefont{Haucke}},
  \bibinfo{author}{\bibfnamefont{S.}~\bibnamefont{Washburn}},
  \bibinfo{author}{\bibfnamefont{A.~D.} \bibnamefont{Benoit}},
  \bibinfo{author}{\bibfnamefont{C.~P.} \bibnamefont{Umbach}},
  \bibnamefont{and} \bibinfo{author}{\bibfnamefont{R.~A.} \bibnamefont{Webb}},
  \bibinfo{journal}{Phys.\ Rev. B} \textbf{\bibinfo{volume}{41}},
  \bibinfo{pages}{12454} (\bibinfo{year}{1990}).

\bibitem[{\citenamefont{Jedema et~al.}(2002)\citenamefont{Jedema, Heersche,
  Filip, Baselmans, and Wees}}]{Jedema_Nature02}
\bibinfo{author}{\bibfnamefont{F.~J.} \bibnamefont{Jedema}},
  \bibinfo{author}{\bibfnamefont{H.~B.} \bibnamefont{Heersche}},
  \bibinfo{author}{\bibfnamefont{A.~T.} \bibnamefont{Filip}},
  \bibinfo{author}{\bibfnamefont{J.~J.~A.} \bibnamefont{Baselmans}},
  \bibnamefont{and} \bibinfo{author}{\bibfnamefont{B.~J.~v.}
  \bibnamefont{Wees}}, \bibinfo{journal}{Nature}
  \textbf{\bibinfo{volume}{416}}, \bibinfo{pages}{713} (\bibinfo{year}{2002}).

\bibitem[{\citenamefont{Lou et~al.}(2006)\citenamefont{Lou, Adelmann, Furis,
  Crooker, Palmstrøm, and Crowell}}]{Lou_PRL06}
\bibinfo{author}{\bibfnamefont{X.}~\bibnamefont{Lou}},
  \bibinfo{author}{\bibfnamefont{C.}~\bibnamefont{Adelmann}},
  \bibinfo{author}{\bibfnamefont{M.}~\bibnamefont{Furis}},
  \bibinfo{author}{\bibfnamefont{S.~A.} \bibnamefont{Crooker}},
  \bibinfo{author}{\bibfnamefont{C.~J.} \bibnamefont{Palmstrøm}},
  \bibnamefont{and} \bibinfo{author}{\bibfnamefont{P.~A.}
  \bibnamefont{Crowell}}, \bibinfo{journal}{Phys.\ Rev.\ Lett.}
  \textbf{\bibinfo{volume}{96}}, \bibinfo{pages}{176603}
  (\bibinfo{year}{2006}).

\bibitem[{\citenamefont{Li et~al.}(2011)\citenamefont{Li, {O. M. J van 't
  Erve}, and Jonker}}]{Li_NC11}
\bibinfo{author}{\bibfnamefont{C.~H.} \bibnamefont{Li}},
  \bibinfo{author}{\bibnamefont{{O. M. J van 't Erve}}}, \bibnamefont{and}
  \bibinfo{author}{\bibfnamefont{B.~T.} \bibnamefont{Jonker}},
  \bibinfo{journal}{Nature Communications} \textbf{\bibinfo{volume}{2}},
  \bibinfo{pages}{245} (\bibinfo{year}{2011}).

\bibitem[{\citenamefont{D'yakonov and Perel'}(1971)}]{Dyakonov_JETP71}
\bibinfo{author}{\bibfnamefont{M.~I.} \bibnamefont{D'yakonov}}
  \bibnamefont{and} \bibinfo{author}{\bibfnamefont{V.~I.}
  \bibnamefont{Perel'}}, \bibinfo{journal}{{JETP} Letters}
  \textbf{\bibinfo{volume}{13}}, \bibinfo{pages}{467} (\bibinfo{year}{1971}).

\bibitem[{\citenamefont{Hirsch}(1999)}]{Hirsch_PRL99}
\bibinfo{author}{\bibfnamefont{J.~E.} \bibnamefont{Hirsch}},
  \bibinfo{journal}{Phys.\ Rev.\ Lett.} \textbf{\bibinfo{volume}{83}},
  \bibinfo{pages}{1834} (\bibinfo{year}{1999}).

\bibitem[{\citenamefont{Hankiewicz et~al.}(2004)\citenamefont{Hankiewicz,
  Molenkamp, Jungwirth, and Sinova}}]{Hankiewicz_PRB04}
\bibinfo{author}{\bibfnamefont{E.~M.} \bibnamefont{Hankiewicz}},
  \bibinfo{author}{\bibfnamefont{L.~W.} \bibnamefont{Molenkamp}},
  \bibinfo{author}{\bibfnamefont{T.}~\bibnamefont{Jungwirth}},
  \bibnamefont{and} \bibinfo{author}{\bibfnamefont{J.}~\bibnamefont{Sinova}},
  \bibinfo{journal}{Phys.\ Rev. B} \textbf{\bibinfo{volume}{70}},
  \bibinfo{pages}{241301} (\bibinfo{year}{2004}).

\bibitem[{\citenamefont{Abanin et~al.}(2009)\citenamefont{Abanin, Shytov,
  Levitov, and Halperin}}]{Abanin_PRB09}
\bibinfo{author}{\bibfnamefont{D.~A.} \bibnamefont{Abanin}},
  \bibinfo{author}{\bibfnamefont{A.~V.} \bibnamefont{Shytov}},
  \bibinfo{author}{\bibfnamefont{L.~S.} \bibnamefont{Levitov}},
  \bibnamefont{and} \bibinfo{author}{\bibfnamefont{B.~I.}
  \bibnamefont{Halperin}}, \bibinfo{journal}{Phys.\ Rev. B}
  \textbf{\bibinfo{volume}{79}}, \bibinfo{pages}{035304}
  (\bibinfo{year}{2009}).

\bibitem[{\citenamefont{Br{\"u}ne et~al.}(2010)\citenamefont{Br{\"u}ne, Roth,
  Novik, K{\"o}nig, Buhmann, Hankiewicz, Hanke, Sinova, and
  Molenkamp}}]{Brune_NP10}
\bibinfo{author}{\bibfnamefont{C.}~\bibnamefont{Br{\"u}ne}},
  \bibinfo{author}{\bibfnamefont{A.}~\bibnamefont{Roth}},
  \bibinfo{author}{\bibfnamefont{E.~G.} \bibnamefont{Novik}},
  \bibinfo{author}{\bibfnamefont{M.}~\bibnamefont{K{\"o}nig}},
  \bibinfo{author}{\bibfnamefont{H.}~\bibnamefont{Buhmann}},
  \bibinfo{author}{\bibfnamefont{E.~M.} \bibnamefont{Hankiewicz}},
  \bibinfo{author}{\bibfnamefont{W.}~\bibnamefont{Hanke}},
  \bibinfo{author}{\bibfnamefont{J.}~\bibnamefont{Sinova}}, \bibnamefont{and}
  \bibinfo{author}{\bibfnamefont{L.}~\bibnamefont{Molenkamp}},
  \bibinfo{journal}{Nat. Phys} \textbf{\bibinfo{volume}{6}},
  \bibinfo{pages}{448} (\bibinfo{year}{2010}).

\bibitem[{Pra()}]{Prange_QHE}
\bibinfo{note}{See e.g.~\emph{The Quantum Hall Effect}, 2$^{nd}$ ed., edited by
  R.E. Prange and S. M. Girvin (Springer, New York, 1990).}

\bibitem[{\citenamefont{Halperin}(1982)}]{Halperin_PRB82}
\bibinfo{author}{\bibfnamefont{B.~I.} \bibnamefont{Halperin}},
  \bibinfo{journal}{Phys.\ Rev. B} \textbf{\bibinfo{volume}{25}},
  \bibinfo{pages}{2185} (\bibinfo{year}{1982}).

\bibitem[{\citenamefont{{MacDonald} and Streda}(1984)}]{MacDonald_PRB84}
\bibinfo{author}{\bibfnamefont{A.~H.} \bibnamefont{{MacDonald}}}
  \bibnamefont{and} \bibinfo{author}{\bibfnamefont{P.}~\bibnamefont{Streda}},
  \bibinfo{journal}{Phys.\ Rev. B} \textbf{\bibinfo{volume}{29}},
  \bibinfo{pages}{1616} (\bibinfo{year}{1984}).

\bibitem[{\citenamefont{{McEuen} et~al.}(1990)\citenamefont{{McEuen}, Szafer,
  Richter, Alphenaar, Jain, Stone, Wheeler, and Sacks}}]{McEuen_PRL90}
\bibinfo{author}{\bibfnamefont{P.~L.} \bibnamefont{{McEuen}}},
  \bibinfo{author}{\bibfnamefont{A.}~\bibnamefont{Szafer}},
  \bibinfo{author}{\bibfnamefont{C.~A.} \bibnamefont{Richter}},
  \bibinfo{author}{\bibfnamefont{B.~W.} \bibnamefont{Alphenaar}},
  \bibinfo{author}{\bibfnamefont{J.~K.} \bibnamefont{Jain}},
  \bibinfo{author}{\bibfnamefont{A.~D.} \bibnamefont{Stone}},
  \bibinfo{author}{\bibfnamefont{R.~G.} \bibnamefont{Wheeler}},
  \bibnamefont{and} \bibinfo{author}{\bibfnamefont{R.~N.} \bibnamefont{Sacks}},
  \bibinfo{journal}{Phys.\ Rev.\ Lett.} \textbf{\bibinfo{volume}{64}},
  \bibinfo{pages}{2062} (\bibinfo{year}{1990}).

\bibitem[{\citenamefont{Takaoka et~al.}(1991)\citenamefont{Takaoka, Sawasaki,
  Tsukagoshi, Oto, Murase, Gamo, and Namba}}]{Takaoka_SSC91}
\bibinfo{author}{\bibfnamefont{S.}~\bibnamefont{Takaoka}},
  \bibinfo{author}{\bibfnamefont{T.}~\bibnamefont{Sawasaki}},
  \bibinfo{author}{\bibfnamefont{K.}~\bibnamefont{Tsukagoshi}},
  \bibinfo{author}{\bibfnamefont{K.}~\bibnamefont{Oto}},
  \bibinfo{author}{\bibfnamefont{K.}~\bibnamefont{Murase}},
  \bibinfo{author}{\bibfnamefont{K.}~\bibnamefont{Gamo}}, \bibnamefont{and}
  \bibinfo{author}{\bibfnamefont{S.}~\bibnamefont{Namba}},
  \bibinfo{journal}{Solid State Communications} \textbf{\bibinfo{volume}{80}},
  \bibinfo{pages}{571} (\bibinfo{year}{1991}).

\bibitem[{\citenamefont{Dolgopolov et~al.}(1991)\citenamefont{Dolgopolov,
  Shashkin, Gusev, and Kvon}}]{Dolgopolov_JETP91}
\bibinfo{author}{\bibfnamefont{V.~T.} \bibnamefont{Dolgopolov}},
  \bibinfo{author}{\bibfnamefont{A.~A.} \bibnamefont{Shashkin}},
  \bibinfo{author}{\bibfnamefont{G.~M.} \bibnamefont{Gusev}}, \bibnamefont{and}
  \bibinfo{author}{\bibfnamefont{Z.~D.} \bibnamefont{Kvon}},
  \bibinfo{journal}{{JETP} Letters} \textbf{\bibinfo{volume}{53}},
  \bibinfo{pages}{484} (\bibinfo{year}{1991}).

\bibitem[{\citenamefont{Hasan and Kane}(2010)}]{Hasan_RMP10}
\bibinfo{author}{\bibfnamefont{M.~Z.} \bibnamefont{Hasan}} \bibnamefont{and}
  \bibinfo{author}{\bibfnamefont{C.~L.} \bibnamefont{Kane}},
  \bibinfo{journal}{Rev. Mod. Phys.} \textbf{\bibinfo{volume}{82}},
  \bibinfo{pages}{3045} (\bibinfo{year}{2010}).

\bibitem[{\citenamefont{Kane and Mele}(2005)}]{Kane_Z2_PRL05}
\bibinfo{author}{\bibfnamefont{C.~L.} \bibnamefont{Kane}} \bibnamefont{and}
  \bibinfo{author}{\bibfnamefont{E.~J.} \bibnamefont{Mele}},
  \bibinfo{journal}{Phys.\ Rev.\ Lett.} \textbf{\bibinfo{volume}{95}},
  \bibinfo{pages}{146802} (\bibinfo{year}{2005}).

\bibitem[{\citenamefont{Bernevig et~al.}(2006)\citenamefont{Bernevig, Hughes,
  and Zhang}}]{Bernevig_Science06}
\bibinfo{author}{\bibfnamefont{B.~A.} \bibnamefont{Bernevig}},
  \bibinfo{author}{\bibfnamefont{T.~L.} \bibnamefont{Hughes}},
  \bibnamefont{and} \bibinfo{author}{\bibfnamefont{S.-C.} \bibnamefont{Zhang}},
  \bibinfo{journal}{Science} \textbf{\bibinfo{volume}{314}},
  \bibinfo{pages}{1757} (\bibinfo{year}{2006}).

\bibitem[{\citenamefont{Konig et~al.}(2007)\citenamefont{Konig, Wiedmann,
  Brune, Roth, Buhmann, Molenkamp, Qi, and Zhang}}]{Konig_Science07}
\bibinfo{author}{\bibfnamefont{M.}~\bibnamefont{Konig}},
  \bibinfo{author}{\bibfnamefont{S.}~\bibnamefont{Wiedmann}},
  \bibinfo{author}{\bibfnamefont{C.}~\bibnamefont{Brune}},
  \bibinfo{author}{\bibfnamefont{A.}~\bibnamefont{Roth}},
  \bibinfo{author}{\bibfnamefont{H.}~\bibnamefont{Buhmann}},
  \bibinfo{author}{\bibfnamefont{L.~W.} \bibnamefont{Molenkamp}},
  \bibinfo{author}{\bibfnamefont{X.-L.} \bibnamefont{Qi}}, \bibnamefont{and}
  \bibinfo{author}{\bibfnamefont{S.-C.} \bibnamefont{Zhang}},
  \bibinfo{journal}{Science} \textbf{\bibinfo{volume}{318}},
  \bibinfo{pages}{766} (\bibinfo{year}{2007}).

\bibitem[{\citenamefont{Roth et~al.}(2009)\citenamefont{Roth, Brune, Buhmann,
  Molenkamp, Maciejko, Qi, and Zhang}}]{Roth_Science09}
\bibinfo{author}{\bibfnamefont{A.}~\bibnamefont{Roth}},
  \bibinfo{author}{\bibfnamefont{C.}~\bibnamefont{Brune}},
  \bibinfo{author}{\bibfnamefont{H.}~\bibnamefont{Buhmann}},
  \bibinfo{author}{\bibfnamefont{L.~W.} \bibnamefont{Molenkamp}},
  \bibinfo{author}{\bibfnamefont{J.}~\bibnamefont{Maciejko}},
  \bibinfo{author}{\bibfnamefont{X.-L.} \bibnamefont{Qi}}, \bibnamefont{and}
  \bibinfo{author}{\bibfnamefont{S.-C.} \bibnamefont{Zhang}},
  \bibinfo{journal}{Science} \textbf{\bibinfo{volume}{325}},
  \bibinfo{pages}{294} (\bibinfo{year}{2009}).

\bibitem[{\citenamefont{Gusev et~al.}(2011)\citenamefont{Gusev, Kvon, Shegai,
  Mikhailov, Dvoretsky, and Portal}}]{Gusev_PRB11}
\bibinfo{author}{\bibfnamefont{G.~M.} \bibnamefont{Gusev}},
  \bibinfo{author}{\bibfnamefont{Z.~D.} \bibnamefont{Kvon}},
  \bibinfo{author}{\bibfnamefont{O.~A.} \bibnamefont{Shegai}},
  \bibinfo{author}{\bibfnamefont{N.~N.} \bibnamefont{Mikhailov}},
  \bibinfo{author}{\bibfnamefont{S.~A.} \bibnamefont{Dvoretsky}},
  \bibnamefont{and} \bibinfo{author}{\bibfnamefont{J.~C.}
  \bibnamefont{Portal}}, \bibinfo{journal}{Phys.\ Rev. B}
  \textbf{\bibinfo{volume}{84}}, \bibinfo{pages}{121302}
  (\bibinfo{year}{2011}).

\bibitem[{\citenamefont{Knez et~al.}(2011)\citenamefont{Knez, Du, and
  Sullivan}}]{Knez_PRL11}
\bibinfo{author}{\bibfnamefont{I.}~\bibnamefont{Knez}},
  \bibinfo{author}{\bibfnamefont{R.-R.} \bibnamefont{Du}}, \bibnamefont{and}
  \bibinfo{author}{\bibfnamefont{G.}~\bibnamefont{Sullivan}},
  \bibinfo{journal}{Phys.\ Rev.\ Lett.} \textbf{\bibinfo{volume}{107}},
  \bibinfo{pages}{136603} (\bibinfo{year}{2011}).

\bibitem[{\citenamefont{Hsieh et~al.}(2012)\citenamefont{Hsieh, Lin, Liu, Duan,
  Bansil, and Fu}}]{Hsieh_arXiv12}
\bibinfo{author}{\bibfnamefont{T.~H.} \bibnamefont{Hsieh}},
  \bibinfo{author}{\bibfnamefont{H.}~\bibnamefont{Lin}},
  \bibinfo{author}{\bibfnamefont{J.}~\bibnamefont{Liu}},
  \bibinfo{author}{\bibfnamefont{W.}~\bibnamefont{Duan}},
  \bibinfo{author}{\bibfnamefont{A.}~\bibnamefont{Bansil}}, \bibnamefont{and}
  \bibinfo{author}{\bibfnamefont{L.}~\bibnamefont{Fu}},
  \bibinfo{journal}{arXiv:1202.1003}  (\bibinfo{year}{2012}).

\bibitem[{\citenamefont{Dziawa et~al.}(2012)\citenamefont{Dziawa, Kowalski,
  Dybko, Buczko, Szczerbakow, Szot, {\L}usakowska, Balasubramanian, Wojek,
  Berntsen et~al.}}]{Dziawa_arXiv12}
\bibinfo{author}{\bibfnamefont{P.}~\bibnamefont{Dziawa}},
  \bibinfo{author}{\bibfnamefont{B.~J.} \bibnamefont{Kowalski}},
  \bibinfo{author}{\bibfnamefont{K.}~\bibnamefont{Dybko}},
  \bibinfo{author}{\bibfnamefont{R.}~\bibnamefont{Buczko}},
  \bibinfo{author}{\bibfnamefont{A.}~\bibnamefont{Szczerbakow}},
  \bibinfo{author}{\bibfnamefont{M.}~\bibnamefont{Szot}},
  \bibinfo{author}{\bibfnamefont{E.}~\bibnamefont{{\L}usakowska}},
  \bibinfo{author}{\bibfnamefont{T.}~\bibnamefont{Balasubramanian}},
  \bibinfo{author}{\bibfnamefont{B.~M.} \bibnamefont{Wojek}},
  \bibinfo{author}{\bibfnamefont{M.~H.} \bibnamefont{Berntsen}},
  \bibnamefont{et~al.}, \bibinfo{journal}{arXiv:1206.1705}
  (\bibinfo{year}{2012}).

\bibitem[{\citenamefont{Xu et~al.}(2012)\citenamefont{Xu, Liu, Alidoust, Qian,
  Neupane, Denlinger, Wang, H, Wray, Cava et~al.}}]{Xu_arXiv12}
\bibinfo{author}{\bibfnamefont{S.}~\bibnamefont{Xu}},
  \bibinfo{author}{\bibfnamefont{C.}~\bibnamefont{Liu}},
  \bibinfo{author}{\bibfnamefont{N.}~\bibnamefont{Alidoust}},
  \bibinfo{author}{\bibfnamefont{D.}~\bibnamefont{Qian}},
  \bibinfo{author}{\bibfnamefont{M.}~\bibnamefont{Neupane}},
  \bibinfo{author}{\bibfnamefont{J.~D.} \bibnamefont{Denlinger}},
  \bibinfo{author}{\bibfnamefont{Y.~J.} \bibnamefont{Wang}},
  \bibinfo{author}{\bibfnamefont{H.~L.} \bibnamefont{H}},
  \bibinfo{author}{\bibfnamefont{L.~A.} \bibnamefont{Wray}},
  \bibinfo{author}{\bibfnamefont{R.~J.} \bibnamefont{Cava}},
  \bibnamefont{et~al.}, \bibinfo{journal}{arXiv:1206.2088}
  (\bibinfo{year}{2012}).

\bibitem[{\citenamefont{Fu}(2011)}]{Fu_PRL11}
\bibinfo{author}{\bibfnamefont{L.}~\bibnamefont{Fu}}, \bibinfo{journal}{Phys.\
  Rev.\ Lett.} \textbf{\bibinfo{volume}{106}}, \bibinfo{pages}{106802}
  (\bibinfo{year}{2011}).

\bibitem[{\citenamefont{Buczko and Cywi{\'n}ski}(2012)}]{Buczko_PRB12}
\bibinfo{author}{\bibfnamefont{R.}~\bibnamefont{Buczko}} \bibnamefont{and}
  \bibinfo{author}{\bibfnamefont{L.}~\bibnamefont{Cywi{\'n}ski}},
  \bibinfo{journal}{Phys.\ Rev. B} \textbf{\bibinfo{volume}{85}},
  \bibinfo{pages}{205319} (\bibinfo{year}{2012}).

\bibitem[{\citenamefont{Grabecki et~al.}(1997)\citenamefont{Grabecki,
  Wr{\'o}bel, Dietl, Sawicki, Domaga{\l}a, Sko{\'s}kiewicz, Papis,
  Kami{\'n}ska, Piotrowska, Leszczy{\'n}ski et~al.}}]{Grabecki_SM97}
\bibinfo{author}{\bibfnamefont{G.}~\bibnamefont{Grabecki}},
  \bibinfo{author}{\bibfnamefont{J.}~\bibnamefont{Wr{\'o}bel}},
  \bibinfo{author}{\bibfnamefont{T.}~\bibnamefont{Dietl}},
  \bibinfo{author}{\bibfnamefont{M.}~\bibnamefont{Sawicki}},
  \bibinfo{author}{\bibfnamefont{J.}~\bibnamefont{Domaga{\l}a}},
  \bibinfo{author}{\bibfnamefont{T.}~\bibnamefont{Sko{\'s}kiewicz}},
  \bibinfo{author}{\bibfnamefont{E.}~\bibnamefont{Papis}},
  \bibinfo{author}{\bibfnamefont{E.}~\bibnamefont{Kami{\'n}ska}},
  \bibinfo{author}{\bibfnamefont{A.}~\bibnamefont{Piotrowska}},
  \bibinfo{author}{\bibfnamefont{M.}~\bibnamefont{Leszczy{\'n}ski}},
  \bibnamefont{et~al.}, \bibinfo{journal}{Superlatt.~Microstr.}
  \textbf{\bibinfo{volume}{22}}, \bibinfo{pages}{51} (\bibinfo{year}{1997}).

\bibitem[{\citenamefont{Wr{\'o}bel}(1996)}]{Wrobel_APPA96}
\bibinfo{author}{\bibfnamefont{J.}~\bibnamefont{Wr{\'o}bel}},
  \bibinfo{journal}{Acta Phys.~Polon.~A} \textbf{\bibinfo{volume}{90}},
  \bibinfo{pages}{691} (\bibinfo{year}{1996}).

\bibitem[{\citenamefont{Oswald et~al.}(1997)\citenamefont{Oswald, Span, Homer,
  Heigl, Ganitzer, Maude, and Portal}}]{Oswald_SSC97}
\bibinfo{author}{\bibfnamefont{J.}~\bibnamefont{Oswald}},
  \bibinfo{author}{\bibfnamefont{G.}~\bibnamefont{Span}},
  \bibinfo{author}{\bibfnamefont{A.}~\bibnamefont{Homer}},
  \bibinfo{author}{\bibfnamefont{G.}~\bibnamefont{Heigl}},
  \bibinfo{author}{\bibfnamefont{P.}~\bibnamefont{Ganitzer}},
  \bibinfo{author}{\bibfnamefont{D.}~\bibnamefont{Maude}}, \bibnamefont{and}
  \bibinfo{author}{\bibfnamefont{J.}~\bibnamefont{Portal}},
  \bibinfo{journal}{Solid State Commun.} \textbf{\bibinfo{volume}{102}},
  \bibinfo{pages}{391} (\bibinfo{year}{1997}).

\bibitem[{\citenamefont{Fu and Kane}(2007)}]{Fu_PRB07}
\bibinfo{author}{\bibfnamefont{L.}~\bibnamefont{Fu}} \bibnamefont{and}
  \bibinfo{author}{\bibfnamefont{C.~L.} \bibnamefont{Kane}},
  \bibinfo{journal}{Phys.\ Rev. B} \textbf{\bibinfo{volume}{76}},
  \bibinfo{pages}{045302} (\bibinfo{year}{2007}).

\bibitem[{\citenamefont{Springholz et~al.}(1993)\citenamefont{Springholz,
  Ihninger, Bauer, Olver, Pastalan, Romaine, and Goldberg}}]{Springholz_APL93}
\bibinfo{author}{\bibfnamefont{G.}~\bibnamefont{Springholz}},
  \bibinfo{author}{\bibfnamefont{G.}~\bibnamefont{Ihninger}},
  \bibinfo{author}{\bibfnamefont{G.}~\bibnamefont{Bauer}},
  \bibinfo{author}{\bibfnamefont{M.~M.} \bibnamefont{Olver}},
  \bibinfo{author}{\bibfnamefont{J.~Z.} \bibnamefont{Pastalan}},
  \bibinfo{author}{\bibfnamefont{S.}~\bibnamefont{Romaine}}, \bibnamefont{and}
  \bibinfo{author}{\bibfnamefont{B.~B.} \bibnamefont{Goldberg}},
  \bibinfo{journal}{Appl.\ Phys.\ Lett.} \textbf{\bibinfo{volume}{63}},
  \bibinfo{pages}{2908} (\bibinfo{year}{1993}).

\bibitem[{\citenamefont{Olver et~al.}(1994)\citenamefont{Olver, Pastalan,
  Romaine, Goldberg, Springholz, Ihninger, , and Bauer}}]{Olver_SSC94}
\bibinfo{author}{\bibfnamefont{M.~M.} \bibnamefont{Olver}},
  \bibinfo{author}{\bibfnamefont{J.~Z.} \bibnamefont{Pastalan}},
  \bibinfo{author}{\bibfnamefont{S.~E.} \bibnamefont{Romaine}},
  \bibinfo{author}{\bibfnamefont{B.~B.} \bibnamefont{Goldberg}},
  \bibinfo{author}{\bibfnamefont{G.}~\bibnamefont{Springholz}},
  \bibinfo{author}{\bibfnamefont{G.}~\bibnamefont{Ihninger}}, ,
  \bibnamefont{and} \bibinfo{author}{\bibfnamefont{G.}~\bibnamefont{Bauer}},
  \bibinfo{journal}{Solid State Commun.} \textbf{\bibinfo{volume}{89}},
  \bibinfo{pages}{693} (\bibinfo{year}{1994}).

\bibitem[{\citenamefont{Chitta et~al.}(2005)\citenamefont{Chitta, Desrat,
  Maude, Piot, Oliveira, Rappl, Ueta, and Abramof}}]{Chitta_PRB05}
\bibinfo{author}{\bibfnamefont{V.~A.} \bibnamefont{Chitta}},
  \bibinfo{author}{\bibfnamefont{W.}~\bibnamefont{Desrat}},
  \bibinfo{author}{\bibfnamefont{D.~K.} \bibnamefont{Maude}},
  \bibinfo{author}{\bibfnamefont{B.~A.} \bibnamefont{Piot}},
  \bibinfo{author}{\bibfnamefont{N.~F.} \bibnamefont{Oliveira}},
  \bibinfo{author}{\bibfnamefont{P.~H.~O.} \bibnamefont{Rappl}},
  \bibinfo{author}{\bibfnamefont{A.~Y.} \bibnamefont{Ueta}}, \bibnamefont{and}
  \bibinfo{author}{\bibfnamefont{E.}~\bibnamefont{Abramof}},
  \bibinfo{journal}{Phys.\ Rev. B} \textbf{\bibinfo{volume}{72}},
  \bibinfo{pages}{195326} (\bibinfo{year}{2005}).

\bibitem[{\citenamefont{Grabecki et~al.}(2005)\citenamefont{Grabecki, Wr\'obel,
  Dietl, Janik, Aleszkiewicz, Papis, Kami\'{n}ska, Piotrowska, Springholz, and
  Bauer}}]{Grabecki_PRB05}
\bibinfo{author}{\bibfnamefont{G.}~\bibnamefont{Grabecki}},
  \bibinfo{author}{\bibfnamefont{J.}~\bibnamefont{Wr\'obel}},
  \bibinfo{author}{\bibfnamefont{T.}~\bibnamefont{Dietl}},
  \bibinfo{author}{\bibfnamefont{E.}~\bibnamefont{Janik}},
  \bibinfo{author}{\bibfnamefont{M.}~\bibnamefont{Aleszkiewicz}},
  \bibinfo{author}{\bibfnamefont{E.}~\bibnamefont{Papis}},
  \bibinfo{author}{\bibfnamefont{E.}~\bibnamefont{Kami\'{n}ska}},
  \bibinfo{author}{\bibfnamefont{A.}~\bibnamefont{Piotrowska}},
  \bibinfo{author}{\bibfnamefont{G.}~\bibnamefont{Springholz}},
  \bibnamefont{and} \bibinfo{author}{\bibfnamefont{G.}~\bibnamefont{Bauer}},
  \bibinfo{journal}{Phys.\ Rev. B} \textbf{\bibinfo{volume}{72}},
  \bibinfo{pages}{125332} (\bibinfo{year}{2005}).

\bibitem[{\citenamefont{Grabecki}(2007)}]{GrabeckI_JAP07}
\bibinfo{author}{\bibfnamefont{G.}~\bibnamefont{Grabecki}},
  \bibinfo{journal}{J.\ Appl.\ Phys.} \textbf{\bibinfo{volume}{101}},
  \bibinfo{pages}{081722} (\bibinfo{year}{2007}).

\bibitem[{\citenamefont{Murakami et~al.}(2004)\citenamefont{Murakami, Nagaosa,
  and Zhang}}]{Murakami_PRL04}
\bibinfo{author}{\bibfnamefont{S.}~\bibnamefont{Murakami}},
  \bibinfo{author}{\bibfnamefont{N.}~\bibnamefont{Nagaosa}}, \bibnamefont{and}
  \bibinfo{author}{\bibfnamefont{S.}~\bibnamefont{Zhang}},
  \bibinfo{journal}{Phys.\ Rev.\ Lett.} \textbf{\bibinfo{volume}{93}},
  \bibinfo{pages}{156804} (\bibinfo{year}{2004}).

\bibitem[{\citenamefont{Dyrda{\l} et~al.}(2009)\citenamefont{Dyrda{\l}, Dugaev,
  and Barna\'s}}]{Dyrdal_EPL09}
\bibinfo{author}{\bibfnamefont{A.}~\bibnamefont{Dyrda{\l}}},
  \bibinfo{author}{\bibfnamefont{V.~K.} \bibnamefont{Dugaev}},
  \bibnamefont{and} \bibinfo{author}{\bibfnamefont{J.}~\bibnamefont{Barna\'s}},
  \bibinfo{journal}{EPL} \textbf{\bibinfo{volume}{85}}, \bibinfo{pages}{67004}
  (\bibinfo{year}{2009}).

\bibitem[{PbT()}]{PbTe_review}
\bibinfo{note}{For review of PbTe properties, see J. I. Ravich, B. A. Efimova,
  and I. A. Smirnov, \emph{Semiconducting Lead Compounds} (Plenum, New York,
  1968); G. Nimtz and B. Schlicht, edited by G. Höhler, Springer Tracts in
  Modern Physics, Vol. 98 (Springer-Verlag, Berlin, 1983), p. 1; \emph{Lead
  Chalcogenides Physics and Applications}, edited by D. Khokhlov (Taylor \&
  Francis, London, 2003).}

\bibitem[{\citenamefont{Tung and Cohen}(1969)}]{Tung_PR69}
\bibinfo{author}{\bibfnamefont{Y.~W.} \bibnamefont{Tung}} \bibnamefont{and}
  \bibinfo{author}{\bibfnamefont{M.~L.} \bibnamefont{Cohen}},
  \bibinfo{journal}{Phys.~Rev.} \textbf{\bibinfo{volume}{180}},
  \bibinfo{pages}{823} (\bibinfo{year}{1969}).

\bibitem[{\citenamefont{Schl{\"u}ter et~al.}(1975)\citenamefont{Schl{\"u}ter,
  Martinez, and Cohen}}]{Schluter_PRB75}
\bibinfo{author}{\bibfnamefont{M.}~\bibnamefont{Schl{\"u}ter}},
  \bibinfo{author}{\bibfnamefont{G.}~\bibnamefont{Martinez}}, \bibnamefont{and}
  \bibinfo{author}{\bibfnamefont{M.~L.} \bibnamefont{Cohen}},
  \bibinfo{journal}{Phys.\ Rev. B} \textbf{\bibinfo{volume}{12}},
  \bibinfo{pages}{650} (\bibinfo{year}{1975}).

\bibitem[{\citenamefont{Gao and Daw}(2008)}]{Gao_PRB08}
\bibinfo{author}{\bibfnamefont{X.}~\bibnamefont{Gao}} \bibnamefont{and}
  \bibinfo{author}{\bibfnamefont{M.~S.} \bibnamefont{Daw}},
  \bibinfo{journal}{Phys.\ Rev. B} \textbf{\bibinfo{volume}{77}},
  \bibinfo{pages}{033103} (\bibinfo{year}{2008}).

\bibitem[{\citenamefont{{\L}usakowski
  et~al.}(2011)\citenamefont{{\L}usakowski, Bogus{\l}awski, and
  Radzy{\'n}ski}}]{Lusakowski_PRB11}
\bibinfo{author}{\bibfnamefont{A.}~\bibnamefont{{\L}usakowski}},
  \bibinfo{author}{\bibfnamefont{P.}~\bibnamefont{Bogus{\l}awski}},
  \bibnamefont{and}
  \bibinfo{author}{\bibfnamefont{T.}~\bibnamefont{Radzy{\'n}ski}},
  \bibinfo{journal}{Phys.\ Rev. B} \textbf{\bibinfo{volume}{83}},
  \bibinfo{pages}{115206} (\bibinfo{year}{2011}).

\bibitem[{\citenamefont{Br\"une et~al.}(2011)\citenamefont{Br\"une, Liu, Novik,
  Hankiewicz, Buhmann, Chen, Qi, Shen, Zhang, and Molenkamp}}]{Brune_PRL11}
\bibinfo{author}{\bibfnamefont{C.}~\bibnamefont{Br\"une}},
  \bibinfo{author}{\bibfnamefont{C.~X.} \bibnamefont{Liu}},
  \bibinfo{author}{\bibfnamefont{E.~G.} \bibnamefont{Novik}},
  \bibinfo{author}{\bibfnamefont{E.~M.} \bibnamefont{Hankiewicz}},
  \bibinfo{author}{\bibfnamefont{H.}~\bibnamefont{Buhmann}},
  \bibinfo{author}{\bibfnamefont{Y.~L.} \bibnamefont{Chen}},
  \bibinfo{author}{\bibfnamefont{X.~L.} \bibnamefont{Qi}},
  \bibinfo{author}{\bibfnamefont{Z.~X.} \bibnamefont{Shen}},
  \bibinfo{author}{\bibfnamefont{S.~C.} \bibnamefont{Zhang}}, \bibnamefont{and}
  \bibinfo{author}{\bibfnamefont{L.~W.} \bibnamefont{Molenkamp}},
  \bibinfo{journal}{Phys.\ Rev.\ Lett.} \textbf{\bibinfo{volume}{106}},
  \bibinfo{pages}{126803} (\bibinfo{year}{2011}).

\bibitem[{Bau()}]{Bauer_ASSP83}
\bibinfo{note}{See, e.g., G. Bauer, W. Jantsch, and E. Bangert, in
  \emph{Advances in Solid State Physics}, edited by P. Grosse (Vieweg,
  Wiesbaden, 1983), Vol. XXIII, p. 27.}

\bibitem[{\citenamefont{Dietl}(1978)}]{Dietl_JP78}
\bibinfo{author}{\bibfnamefont{T.}~\bibnamefont{Dietl}}, \bibinfo{journal}{J.
  Phys. (Paris)} \textbf{\bibinfo{volume}{39-C6}}, \bibinfo{pages}{1081}
  (\bibinfo{year}{1978}).

\bibitem[{\citenamefont{Jedrzejczak et~al.}(1978)\citenamefont{Jedrzejczak,
  Guillot, and Martinez}}]{Jedrzejczak_PRB78}
\bibinfo{author}{\bibfnamefont{A.}~\bibnamefont{Jedrzejczak}},
  \bibinfo{author}{\bibfnamefont{D.}~\bibnamefont{Guillot}}, \bibnamefont{and}
  \bibinfo{author}{\bibfnamefont{G.}~\bibnamefont{Martinez}},
  \bibinfo{journal}{Phys.\ Rev. B} \textbf{\bibinfo{volume}{17}},
  \bibinfo{pages}{829} (\bibinfo{year}{1978}).

\bibitem[{\citenamefont{Prinz et~al.}(1999)\citenamefont{Prinz, Brunthaler,
  Ueta, Springholz, Bauer, Grabecki, and Dietl}}]{Prinz_PRB99}
\bibinfo{author}{\bibfnamefont{A.}~\bibnamefont{Prinz}},
  \bibinfo{author}{\bibfnamefont{G.}~\bibnamefont{Brunthaler}},
  \bibinfo{author}{\bibfnamefont{Y.}~\bibnamefont{Ueta}},
  \bibinfo{author}{\bibfnamefont{G.}~\bibnamefont{Springholz}},
  \bibinfo{author}{\bibfnamefont{G.}~\bibnamefont{Bauer}},
  \bibinfo{author}{\bibfnamefont{G.}~\bibnamefont{Grabecki}}, \bibnamefont{and}
  \bibinfo{author}{\bibfnamefont{T.}~\bibnamefont{Dietl}},
  \bibinfo{journal}{Phys.\ Rev. B} \textbf{\bibinfo{volume}{59}},
  \bibinfo{pages}{12983} (\bibinfo{year}{1999}).

\bibitem[{\citenamefont{Kaufmann}(1984)}]{Kaufmann_PRB84}
\bibinfo{author}{\bibfnamefont{B.}~\bibnamefont{Kaufmann}},
  \bibinfo{journal}{Phys.\ Rev. B} \textbf{\bibinfo{volume}{30}},
  \bibinfo{pages}{4640} (\bibinfo{year}{1984}).

\bibitem[{Spr()}]{Springholz_03}
\bibinfo{note}{G. Springholz, in \emph{Lead Chalcogenides Physics and
  Applications}, edited by D. Khokhlov (Taylor \& Francis, London 2003), p.
  123.}

\bibitem[{\citenamefont{Springholz et~al.}(1996)\citenamefont{Springholz, Ueta,
  Frank, and Bauer}}]{Springholz_APL96}
\bibinfo{author}{\bibfnamefont{G.}~\bibnamefont{Springholz}},
  \bibinfo{author}{\bibfnamefont{A.~Y.} \bibnamefont{Ueta}},
  \bibinfo{author}{\bibfnamefont{N.}~\bibnamefont{Frank}}, \bibnamefont{and}
  \bibinfo{author}{\bibfnamefont{G.}~\bibnamefont{Bauer}},
  \bibinfo{journal}{Appl.\ Phys.\ Lett.} \textbf{\bibinfo{volume}{69}},
  \bibinfo{pages}{2822} (\bibinfo{year}{1996}).

\bibitem[{\citenamefont{Ueta et~al.}(1997)\citenamefont{Ueta, Springholz, and
  Bauer}}]{Ueta_JCG97}
\bibinfo{author}{\bibfnamefont{A.}~\bibnamefont{Ueta}},
  \bibinfo{author}{\bibfnamefont{G.}~\bibnamefont{Springholz}},
  \bibnamefont{and} \bibinfo{author}{\bibfnamefont{G.}~\bibnamefont{Bauer}},
  \bibinfo{journal}{J.~Cryst.~Growth} \textbf{\bibinfo{volume}{175/176}},
  \bibinfo{pages}{1022} (\bibinfo{year}{1997}).

\bibitem[{Tra()}]{Tranta88}
\bibinfo{note}{B.~Tranta and H.~Clemens, in \emph{New Developments in
  Semiconductor Physics}, Lecture Notes in Physics Vol.~301, edited by
  G.~Ferenczi and F.~Beleznay (Springer Verlag, Berlin, 1988), p. 281.}

\bibitem[{\citenamefont{Grabecki et~al.}(2004)\citenamefont{Grabecki,
  Wr{\'o}bel, Dietl, Papis, Kami{\'n}ska, Piotrowska, Ratuszna, Springholz, ,
  and Bauer}}]{Grabecki_PE04}
\bibinfo{author}{\bibfnamefont{G.}~\bibnamefont{Grabecki}},
  \bibinfo{author}{\bibfnamefont{J.}~\bibnamefont{Wr{\'o}bel}},
  \bibinfo{author}{\bibfnamefont{T.}~\bibnamefont{Dietl}},
  \bibinfo{author}{\bibfnamefont{E.}~\bibnamefont{Papis}},
  \bibinfo{author}{\bibfnamefont{E.}~\bibnamefont{Kami{\'n}ska}},
  \bibinfo{author}{\bibfnamefont{A.}~\bibnamefont{Piotrowska}},
  \bibinfo{author}{\bibfnamefont{A.}~\bibnamefont{Ratuszna}},
  \bibinfo{author}{\bibfnamefont{G.}~\bibnamefont{Springholz}}, ,
  \bibnamefont{and} \bibinfo{author}{\bibfnamefont{G.}~\bibnamefont{Bauer}},
  \bibinfo{journal}{Physica E (Amsterdam)} \textbf{\bibinfo{volume}{20}},
  \bibinfo{pages}{236} (\bibinfo{year}{2004}).

\bibitem[{\citenamefont{Zogg et~al.}(1994)\citenamefont{Zogg, Blunier, Fach,
  Maissen, Müller, Teodoropol, Meyer, Kostorz, Dommann, and
  Richmond}}]{Zogg_PRB94}
\bibinfo{author}{\bibfnamefont{H.}~\bibnamefont{Zogg}},
  \bibinfo{author}{\bibfnamefont{S.}~\bibnamefont{Blunier}},
  \bibinfo{author}{\bibfnamefont{A.}~\bibnamefont{Fach}},
  \bibinfo{author}{\bibfnamefont{C.}~\bibnamefont{Maissen}},
  \bibinfo{author}{\bibfnamefont{P.}~\bibnamefont{Müller}},
  \bibinfo{author}{\bibfnamefont{S.}~\bibnamefont{Teodoropol}},
  \bibinfo{author}{\bibfnamefont{V.}~\bibnamefont{Meyer}},
  \bibinfo{author}{\bibfnamefont{G.}~\bibnamefont{Kostorz}},
  \bibinfo{author}{\bibfnamefont{A.}~\bibnamefont{Dommann}}, \bibnamefont{and}
  \bibinfo{author}{\bibfnamefont{T.}~\bibnamefont{Richmond}},
  \bibinfo{journal}{Phys.\ Rev. B} \textbf{\bibinfo{volume}{50}},
  \bibinfo{pages}{10801} (\bibinfo{year}{1994}).

\bibitem[{\citenamefont{Grabecki et~al.}(1995)\citenamefont{Grabecki, Takeyama,
  Adachi, Takagi, Dietl, Kami{\'n}ska, Piotrowska, Papis, Frank, and
  Bauer}}]{Grabecki_JJAP95}
\bibinfo{author}{\bibfnamefont{G.}~\bibnamefont{Grabecki}},
  \bibinfo{author}{\bibfnamefont{S.}~\bibnamefont{Takeyama}},
  \bibinfo{author}{\bibfnamefont{S.}~\bibnamefont{Adachi}},
  \bibinfo{author}{\bibfnamefont{Y.}~\bibnamefont{Takagi}},
  \bibinfo{author}{\bibfnamefont{T.}~\bibnamefont{Dietl}},
  \bibinfo{author}{\bibfnamefont{E.}~\bibnamefont{Kami{\'n}ska}},
  \bibinfo{author}{\bibfnamefont{A.}~\bibnamefont{Piotrowska}},
  \bibinfo{author}{\bibfnamefont{E.}~\bibnamefont{Papis}},
  \bibinfo{author}{\bibfnamefont{N.}~\bibnamefont{Frank}}, \bibnamefont{and}
  \bibinfo{author}{\bibfnamefont{G.}~\bibnamefont{Bauer}},
  \bibinfo{journal}{Jpn.~J.~Appl.~Phys.} \textbf{\bibinfo{volume}{34}},
  \bibinfo{pages}{4433} (\bibinfo{year}{1995}).

\bibitem[{\citenamefont{Al'tshuler and Shklovski}(1986)}]{Altshuler_JETP86}
\bibinfo{author}{\bibfnamefont{B.~L.} \bibnamefont{Al'tshuler}}
  \bibnamefont{and} \bibinfo{author}{\bibfnamefont{B.~I.}
  \bibnamefont{Shklovski}}, \bibinfo{journal}{Sov.~Phys.~JETP}
  \textbf{\bibinfo{volume}{64}}, \bibinfo{pages}{127} (\bibinfo{year}{1986}).

\bibitem[{\citenamefont{Lee et~al.}(1987)\citenamefont{Lee, Stone, and
  Fukuyama}}]{Lee_PRB87}
\bibinfo{author}{\bibfnamefont{P.~A.} \bibnamefont{Lee}},
  \bibinfo{author}{\bibfnamefont{A.~D.} \bibnamefont{Stone}}, \bibnamefont{and}
  \bibinfo{author}{\bibfnamefont{H.}~\bibnamefont{Fukuyama}},
  \bibinfo{journal}{Phys.\ Rev. B} \textbf{\bibinfo{volume}{35}},
  \bibinfo{pages}{1039} (\bibinfo{year}{1987}).

\bibitem[{\citenamefont{Rogalski}(2000)}]{Rogalski}
\bibinfo{author}{\bibfnamefont{A.}~\bibnamefont{Rogalski}},
  \emph{\bibinfo{title}{Infrared Detectors}} (\bibinfo{publisher}{Gordon \&
  Breach}, \bibinfo{year}{2000}).

\bibitem[{\citenamefont{Zimin et~al.}(2002)\citenamefont{Zimin, Alchalabi, and
  Zogg}}]{Zimin_PE02}
\bibinfo{author}{\bibfnamefont{D.}~\bibnamefont{Zimin}},
  \bibinfo{author}{\bibfnamefont{K.}~\bibnamefont{Alchalabi}},
  \bibnamefont{and} \bibinfo{author}{\bibfnamefont{H.}~\bibnamefont{Zogg}},
  \bibinfo{journal}{Physica E} \textbf{\bibinfo{volume}{13}},
  \bibinfo{pages}{1220} (\bibinfo{year}{2002}).

\bibitem[{\citenamefont{Chang et~al.}(1980)\citenamefont{Chang, Singer, and
  Northrop}}]{Chang_JPD80}
\bibinfo{author}{\bibfnamefont{B.}~\bibnamefont{Chang}},
  \bibinfo{author}{\bibfnamefont{K.~E.} \bibnamefont{Singer}},
  \bibnamefont{and} \bibinfo{author}{\bibfnamefont{D.~C.}
  \bibnamefont{Northrop}}, \bibinfo{journal}{J.~Phys.~D: Appl.~Phys.}
  \textbf{\bibinfo{volume}{13}}, \bibinfo{pages}{715} (\bibinfo{year}{1980}).

\bibitem[{\citenamefont{Grabecki et~al.}(2010)\citenamefont{Grabecki, Kolwas,
  Wr{\'o}bel, Kapcia, Pu{\'z}niak, Jakie{\l}a, Aleszkiewicz, Dietl, Springholz,
  and Bauer}}]{Grabecki_JAP10}
\bibinfo{author}{\bibfnamefont{G.}~\bibnamefont{Grabecki}},
  \bibinfo{author}{\bibfnamefont{K.~A.} \bibnamefont{Kolwas}},
  \bibinfo{author}{\bibfnamefont{J.}~\bibnamefont{Wr{\'o}bel}},
  \bibinfo{author}{\bibfnamefont{K.}~\bibnamefont{Kapcia}},
  \bibinfo{author}{\bibfnamefont{R.}~\bibnamefont{Pu{\'z}niak}},
  \bibinfo{author}{\bibfnamefont{R.}~\bibnamefont{Jakie{\l}a}},
  \bibinfo{author}{\bibfnamefont{M.}~\bibnamefont{Aleszkiewicz}},
  \bibinfo{author}{\bibfnamefont{T.}~\bibnamefont{Dietl}},
  \bibinfo{author}{\bibfnamefont{G.}~\bibnamefont{Springholz}},
  \bibnamefont{and} \bibinfo{author}{\bibfnamefont{G.}~\bibnamefont{Bauer}},
  \bibinfo{journal}{J.\ Appl.\ Phys.} \textbf{\bibinfo{volume}{108}},
  \bibinfo{pages}{053714} (\bibinfo{year}{2010}).

\bibitem[{Bee()}]{Beenakker}
\bibinfo{note}{C.~W.~J. Beenakker and H.~van Houten, \emph{Solid State
  Physics}, edited by H. Ehrenreich and D. Turnbull (Academic, New York, 1991),
  Vol. 44, p. 1.}

\bibitem[{\citenamefont{Pannetier et~al.}(1984)\citenamefont{Pannetier,
  Chaussy, Rammal, and Gandit}}]{Pannetier_PRL84}
\bibinfo{author}{\bibfnamefont{B.}~\bibnamefont{Pannetier}},
  \bibinfo{author}{\bibfnamefont{J.}~\bibnamefont{Chaussy}},
  \bibinfo{author}{\bibfnamefont{R.}~\bibnamefont{Rammal}}, \bibnamefont{and}
  \bibinfo{author}{\bibfnamefont{P.}~\bibnamefont{Gandit}},
  \bibinfo{journal}{Phys.\ Rev.\ Lett.} \textbf{\bibinfo{volume}{53}},
  \bibinfo{pages}{718} (\bibinfo{year}{1984}).

\bibitem[{\citenamefont{Figielski et~al.}(2000)\citenamefont{Figielski,
  Wosi{\'n}ski, and Makosa}}]{Figielski_pssb00}
\bibinfo{author}{\bibfnamefont{T.}~\bibnamefont{Figielski}},
  \bibinfo{author}{\bibfnamefont{T.}~\bibnamefont{Wosi{\'n}ski}},
  \bibnamefont{and} \bibinfo{author}{\bibfnamefont{A.}~\bibnamefont{Makosa}},
  \bibinfo{journal}{Phys.~Status Solidi B} \textbf{\bibinfo{volume}{222}},
  \bibinfo{pages}{151} (\bibinfo{year}{2000}).

\bibitem[{\citenamefont{Yuan et~al.}(1993)\citenamefont{Yuan, Krenn,
  Springholz, and Bauer}}]{Yuan_PRB93}
\bibinfo{author}{\bibfnamefont{S.}~\bibnamefont{Yuan}},
  \bibinfo{author}{\bibfnamefont{H.}~\bibnamefont{Krenn}},
  \bibinfo{author}{\bibfnamefont{G.}~\bibnamefont{Springholz}},
  \bibnamefont{and} \bibinfo{author}{\bibfnamefont{G.}~\bibnamefont{Bauer}},
  \bibinfo{journal}{Phys.\ Rev. B} \textbf{\bibinfo{volume}{47}},
  \bibinfo{pages}{7213} (\bibinfo{year}{1993}).

\end{thebibliography}

\end{document}